\documentclass[aps,prx,onecolumn,superscriptaddress,showpacs,amsmath,amssymb,footinbib,notitlepage,longbibliography]{revtex4-1}

\usepackage{mathtools} \usepackage{amsfonts} \usepackage{amssymb}
\usepackage{graphicx} \usepackage{color}

\usepackage{bm} \usepackage{braket} \usepackage{gensymb}

\usepackage{hyperref}

\begin{document}

\title{On the general properties of non-linear optical conductivities}

\author{Haruki Watanabe} \author{Yankang Liu} \affiliation{Department of
Applied Physics, University of Tokyo, Tokyo 113-8656, Japan.}

\author{Masaki Oshikawa} \affiliation{Institute for Solid State Physics,
University of Tokyo, Kashiwa 277-8581, Japan.}  \affiliation{Kavli
Institute for the Physics and Mathematics of the Universe (WPI),
University of Tokyo, Kashiwa 277-8583, Japan}

\begin{abstract}
The optical conductivity is the basic defining property of materials
characterizing the current response toward time-dependent electric
fields.  In this work, following the approach of Kubo's response
theory, we study the general properties of the nonlinear optical
conductivities of quantum many-body systems both in equilibrium and
non-equilibrium.  We obtain an expression of the second- and the
third-order optical conductivity in terms of correlation functions and
present a perturbative proof of the generalized Kohn formula proposed
recently.  We also discuss a generalization of the $f$-sum rule to a
non-equilibrium setting by focusing on the instantaneous response.
\end{abstract}

\maketitle

\section{Introduction}

The electric conductivity describes the response of the current density
$j_i(t)$ toward a time-dependent electric field $E_j(t)$.  In the
Fourier space, the linear optical conductivity $\sigma_i^j(\omega)$
($i,j$ are the spatial indices) is the proportionality constant
connecting $j_i(\omega)$ to $E_i(\omega)$:
\begin{align}
j_i(\omega)=\sum_j\sigma_i^j(\omega)E_j(\omega)+O(E^2).
\end{align}
There has been a long history of studies on the general properties of
$\sigma_i^j(\omega)$. (See Ref.~\cite{Resta_2018} and the references therein.) For example, the optical conductivity obeys the
frequency-sum rule ($f$-sum rule)~\cite{Resta_2018}; that is, the
integral
\begin{align}
\int_{-\infty}^{\infty}d\omega\sigma_i^{j}(\omega)~\label{fsumrule}
\end{align}
is solely determined by an expectation value in the absence of the
electric field.  Furthermore, the optical conductivity is known to have
the following generic structure:
\begin{align}
\sigma_{i}^{j}(\omega)=\frac{i}{\omega+i\eta}\mathcal{D}_{i}^{j}+\sigma_{i\,(\text{regular})}^{j}(\omega),\label{standardform}
\end{align}
where $\mathcal{D}_i^j$ is called the Drude weight that characterizes
the singular part of $\sigma_i^j(\omega)$ around $\omega=0$ and
$\sigma_{i\,(\text{regular})}^{j}(\omega)$ is the regular part that
includes all other terms.  The Drude weight is a useful measure
distinguishing ideal conductors from insulators and non-ideal
conductors~\cite{SWM2000}. More than a half century ago,
Kohn~\cite{Kohn1964} showed that the Drude weight at zero temperature is given by the curvature of the ground
state energy $\mathcal{E}_0(\vec{A})$ as a function of the vector
potential $\vec{A}$,
\begin{align}
&\mathcal{D}_i^j=\frac{1}{V}\frac{\partial^{2}\mathcal{E}_0(\vec{A})}{\partial
A_i\partial A_j}\Big|_{\vec{A}=0}.\label{Kohn1}
\end{align}
This is nowadays known as the Kohn formula~\cite{Resta_2018}. An
extension to a finite temperature was achieved in
Ref.~\cite{PhysRevLett.74.972}.

Recently, two of us proposed~\cite{2003.10390} a generalization of the $f$-sum
rule and the Kohn formula to the $N$-th order optical conductivity
$\sigma_{i}^{i_1\dots i_N}(\omega_1,\dots,\omega_N)$ [defined by
Eqs.~\eqref{defsigma} and \eqref{deffourier1}] through a heuristic argument
utilizing extreme quantum processes in the quench or the adiabatic
limit.  Historically, however, the $f$-sum rule and the Kohn formula for
the linear optical conductivity were derived using the concrete
expression of $\sigma_i^j(\omega)$ in terms of current-current
correlation functions obtained from the linear response
theory~\cite{Kubo} as we review in
Sec.~\ref{Kohn}~\cite{Kohn1964,PhysRevLett.74.972,Resta_2018}.  The main
result of this work is to put forward this analysis to higher-order
conductivities and provide a proof of the generalized Kohn formula [in
Eq.~\eqref{main2} below] via the perturbation theory for the second- and
third-order response.

On the way to achieve this goal, we also find it
possible to further generalize the $f$-sum rule to arbitrary non-equilibrium states by focusing on the \emph{instantaneous} response [see
Eq.~\eqref{main1}], which may be seen as another result of this
work. This idea was briefly sketched in Ref.~\cite{2003.10390} without concrete formulation. Our study extends earlier works~\cite{Shimizu,ShimizuYuge1,ShimizuYuge2,RevModPhys.86.779} on the $f$-sum rule in several respects.
\begin{itemize}

\item In Ref.~\cite{Shimizu}, the $f$-sum rules were formulated for general nonlinear
response functions at higher orders. Since the electric conductivity
represents the response of the electric current to the applied
electric field, the nonlinear $f$-sum rules discussed in this paper
might appear to be a special case of such a general result. However,
the formulation in Ref.~\cite{Shimizu} by itself does not directly give the $f$-sum
rules for the optical conductivity, because the perturbed
Hamiltonian in our problem is written in
terms of the vector potential $\vec{A}(t)$, not the electric field $\vec{E}(t)$ for which
the optical conductivity is defined.

\item In Refs.~\cite{Shimizu,ShimizuYuge1}, the electric field was described in terms of the scalar potential $\phi=-\vec{E}(t)\cdot\sum_j\hat{\vec{q}}^j$ (i.e., the length gauge~\cite{Parker2019}). This cannot represent a uniform electric field under the periodic boundary condition because the position operator becomes ill-defined~\cite{PhysRevLett.80.1800}. Thus the $f$-sum rules for a uniform electric field in a finite system with the periodic boundary condition cannot be derived in this way.

\item In Refs.~\cite{Shimizu,ShimizuYuge1,ShimizuYuge2} the $f$-sum rule was derived for, in addition to the equilibrium states, non-equilibrium steady states or non-equilibrium states prepared in a specific protocol although with a large degree of freedoms including the functional form of the time-dependence of the pump field.

\item In Refs.~\cite{Shimizu,ShimizuYuge1,ShimizuYuge2,RevModPhys.86.779}, the separation of the Hamiltonian into the kinetic term (which is quadratic in particle field operators) and potential terms is assumed.
Furthermore, the Newtonian kinetic energy $\hat{K}=\sum_{j}\hat{\vec{p}}_j^2/2m$ was assumed in Refs.~\cite{Shimizu,ShimizuYuge1}.
While the $f$-sum rule for the nonlinear conductivities was discussed in Ref.~\cite{Shimizu}, the assumed form of the kinetic energy leads to vanishing $f$-sum for nonlinear conductivities at all (second and higher) orders.
On the other hand, in Refs.~\cite{ShimizuYuge2,RevModPhys.86.779}, more general (but still quadratic in field operators) kinetic term is considered, but the $f$-sum rule was derived only for the linear optical conductivity.

\end{itemize}

In contrast, we derive the compact explicit form of the $f$-sum rule of optical conductivities at arbitrary orders for most general non-equilibrium states, without assuming any specific form of the Hamiltonian.
In particular, we find non-trivial $f$-sum rules for the second and the higher order non-linear conductivities for general Hamiltonians beyond the Newtonian or quadratic kinetic term. Wherever overlaps, our result is consistent with the earlier works.
Our analysis also clarifies that the $f$-sum rule is not a property specific to  the class of non-equilibrium states which were studied in Refs.~\cite{Shimizu,ShimizuYuge1,ShimizuYuge2}, but essentially a statement on the instantaneous response of any state (although the standard representation in terms of the frequency integral is applicable only to stationary states).
We also describe the electric field by the time-dependent gauge field (i.e., the velocity gauge~\cite{Parker2019}), which can naturally represent a uniform electric field under any boundary condition including the periodic one.
For this purpose, we will keep the discussion
applicable to general time-dependent, non-equilibrium states as much as possible.
Furthermore, our analysis of the $f$-sum rules clarifies the similarlty to, and the difference from, the (non-linear) Kohn formulas, which is the other main topic of the present paper.

\section{Setup and Definitions}
\label{setup} Here we explain the setting of our study and give the
definition of the nonlinear optical conductivity and the nonlinear Drude
weight.

\subsection{Setup}
We consider quantum many-body systems defined on the $d$-dimensional
cubic lattice~\footnote{See Ref.~\cite{2003.10390} for the generalization to arbitrary lattice.}.  The system size $V$ is kept finite with an arbitrary
boundary condition.  Let $\hat{H}_{0}(t)$ be the Hamiltonian of the
system, which is allowed to explicitly depend on $t$.  The Hamiltonian
may contain arbitrary forms of kinetic terms and interactions, but all
terms are required to be short-ranged and U(1) symmetric.  If the
initial density matrix is $\hat{\rho}_0$, the density matrix at a later
time $t$ is given by
\begin{equation}
\hat{\rho}_{0}(t)=\hat{S}_0(t)\hat{\rho}_0\hat{S}_0(t)^\dagger,\label{defrho0t}
\end{equation}
where $\hat{S}_0(t)$ is the time-evolution operator
\begin{align}
\hat{S}_0(t)=\mathcal{T}\exp\left(-i
\int_0^tdt'\hat{H}_0(t')\right).\label{defS0}
\end{align}
Since we have not put any restriction on the initial density matrix,
$\hat{\rho}_{0}(t)$ can describe an arbitrary equilibrium and
non-equilibrium state.

We perturb this system by applying a uniform electric field $\vec{E}(t)
\equiv d \vec{A}(t)/dt$ via the vector potential
$\vec{A}(t)=(A_x(t),A_y(t),\dots)$.  (Note that our sign convention of
$\vec{A}(t)$ is opposite to the standard one. ) The perturbed Hamiltonian
is denoted by $\hat{H}(t,\vec{A}(t))$. $\vec{A}(t)$ is set to be
$\vec{0}$ for $t\leq0$ and is turned on continuously at $t=0$.  The word
``perturbation" in this work is exclusively used regarding the external
field; the effect of many-body interactions and disorders, if any, can
be fully taken into account in $\hat{\rho}_0(t)$. We assume that
$\hat{H}(t,\vec{A})$ is analytic with respect to $\vec{A}$ so that
\begin{align}
&\hat{H}_{i_1\dots
i_N}(t)\equiv\frac{\partial^N\hat{H}(t,\vec{A})}{\partial
A_{i_1}\dots\partial A_{i_N}}\Big|_{\vec{A}=\vec{0}}\label{defHN2}
\end{align}
is well-defined for any integer $N\geq1$.

We are interested in the response of the current density toward the
applied electric field. The current density operator is defined by
\begin{equation}
\hat{\vec{j}}(t,\vec{A})\equiv\frac{1}{V}\frac{\partial
\hat{H}(t,\vec{A})}{\partial \vec{A}}.\label{defj}
\end{equation}
If $\hat{\rho}(t)$ is the perturbed density matrix fully including the
effect of $\vec{A}(t)$, the current expectation value at time $t$ reads
\begin{equation}
j_i(t)\equiv\text{tr}\big(\hat{j}_i(t,\vec{A}(t))\hat{\rho}(t)\big)\quad
(i=x,y,\dots).
\end{equation}
The spontaneous current $j_i(t)|_{\vec{A}=0}$ vanishes in the Gibbs states
and in the ground state~\cite{Tada,Bachmann,WatanabeBloch}, while it may
be nonzero in other more general states. The $N$-th order response of
$j_i(t)$ ($N\geq1$) may be written as
\begin{align}
j_{i}^{(N)}(t)&=\frac{1}{N!}\sum_{i_1,\dots,i_N}\int_0^{t+\epsilon}
dt_1\dots \int_0^{t+\epsilon} dt_N \sigma_{i}^{i_1\dots
i_N}(t,t_1,\dots, t_N)\prod_{\ell=1}^NE_{i_\ell}(t_\ell),
\label{defsigma}
\end{align}
which gives the definition of the $N$-th order conductivity. Here, the
summation of $i_\ell$'s ($\ell=1,\dots,N$) is taken over $x,y,\dots$. An
infinitesimal parameter $\epsilon>0$ is included to properly treat
possible $\delta$-functions. As $\sigma_{i}^{i_1\dots i_N}(t,t_1,\dots,
t_N)$ vanishes whenever $t_\ell>t$ for any $\ell=1,\dots,N$ due to the
causality, the value of $\epsilon$ does not affect the integral.  The
nonlinear conductivity $\sigma_{i}^{i_1\dots i_N}(t,t_1,\dots, t_N)$ is
symmetric with respect to the exchange of any pair of $(i_\ell,t_\ell)$
and $(i_{\ell'},t_{\ell'})$.

For our discussions below, we find it useful to introduce another set of
response functions $\phi_{i}^{i_1\dots i_N}(t,t_1,\dots, t_N)$. They are
defined similarly to $\sigma_{i}^{i_1\dots i_N}(t,t_1,\dots, t_N)$ in
Eq.~\eqref{defsigma} but $\vec{E}(t)$ there is replaced with
$\vec{A}(t)$:
\begin{align}
j_{i}^{(N)}(t)&=\frac{1}{N!}\sum_{i_1,\dots,i_N}\int_0^{t+\epsilon}dt_1\dots
\int_0^{t+\epsilon} dt_N \phi_{i}^{i_1\dots i_N}(t,t_1,\dots,
t_N)\prod_{\ell=1}^NA_{i_\ell}(t_\ell).  \label{defg}
\end{align}
In terms of the response function $\phi_{i}^{i_1\dots i_N}(t,t_1,\dots,
t_N)$, the conductivity $\sigma_{i}^{i_1\dots i_N}(t,t_1,\dots, t_N)$ is
given by
\begin{align}
\sigma_{i}^{i_1\dots i_N}(t,t_1,\dots, t_N)=\int_{t_1}^{t+\epsilon}dt_1'
\dots \int_{t_N}^{t+\epsilon}dt_N' \phi_{i}^{i_1\dots i_N}(t,t_1',\dots,
t_N').\label{gtosigma}
\end{align}
This can be seen by expressing $\vec{A}(t)$ as
$\vec{A}(t)=\int_{0}^{t}dt'\vec{E}(t')$ and by performing integration by
parts one by one for $t_1,\dots,t_N$ in Eq.~\eqref{defg}.
\begin{align}
&\int_0^{t+\epsilon} dt_1\dots \int_0^{t+\epsilon} dt_N
\phi_{i}^{i_1\dots i_N}(t,t_1,\dots,
t_N)\prod_{\ell=1}^NA_{i_\ell}(t_\ell)\notag\\ &=\int_0^{t+\epsilon}
dt_1\dots \int_0^{t+\epsilon} dt_N \left(\frac{d}{dt_1}\int_0^{t_1}dt_1'
\phi_{i}^{i_1\dots i_N}(t,t_1',t_2,\dots, t_N)\right)\left(\int_0^{t_1}
dt_1'' E_{i_1}(t_1'')\right)\prod_{\ell=2}^NA_{i_\ell}(t_\ell)\notag\\
&=\int_0^{t+\epsilon} dt_2\dots \int_0^{t+\epsilon} dt_N
\int_0^{{t+\epsilon}}dt_1' \phi_{i}^{i_1\dots i_N}(t,t_1',t_2,\dots,
t_N)\int_0^{{t+\epsilon}} dt_1''
E_{i_1}(t_1'')\prod_{\ell=2}^NA_{i_\ell}(t_\ell)\notag\\
&\quad-\int_0^{t+\epsilon} dt_1\dots \int_0^{t+\epsilon} dt_N
\int_0^{t_1}dt_1' \phi_{i}^{i_1\dots i_N}(t,t_1',t_2,\dots, t_N)
E_{i_1}(t_1)\prod_{\ell=2}^NA_{i_\ell}(t_\ell)\notag\\
&=\int_0^{t+\epsilon} dt_1\dots \int_0^{t+\epsilon} dt_N
\left(\int_{t_1}^{{t+\epsilon}}dt_1' \phi_{i}^{i_1\dots
i_N}(t,t_1',t_2,\dots,
t_N)\right)E_{i_1}(t_1)\prod_{\ell=2}^NA_{i_\ell}(t_\ell)\notag\\
&=\dots\notag\\ &=\int_0^{t+\epsilon} dt_1\dots \int_0^{t+\epsilon} dt_N
\left(\int_{t_1}^{{t+\epsilon}}dt_1'\dots \int_{t_N}^{{t+\epsilon}}dt_N'
\phi_{i}^{i_1\dots i_N}(t,t_1',\dots,
t_N')\right)\prod_{\ell=1}^NE_{i_\ell}(t_\ell).
\end{align}
Comparing the last line with Eq.~\eqref{defsigma}, we obtain
Eq.~\eqref{gtosigma}.

\subsection{Instantaneous response}

Let us define the instantaneous conductivity by
\begin{align}
\mathcal{I}_{i}^{i_1\dots i_N}(t)&\equiv\lim_{t'\rightarrow
t-0}\sigma_{i}^{i_1\dots i_N}(t,t',\dots, t').\label{sigmainst}
\end{align}
In terms of $\phi_{i}^{i_1\dots i_N}(t,t_1,\dots, t_N)$, we have
\begin{align}
\mathcal{I}_{i}^{i_1\dots i_N}(t)=\lim_{t'\rightarrow
t-0}\int_{t'}^{t+\epsilon}dt_1' \dots \int_{t'}^{t+\epsilon}dt_N'
\phi_{i}^{i_1\dots i_N}(t,t_1',\dots, t_N'),
\end{align}
which implies that $\phi_{i}^{i_1\dots i_N}(t,t_1,\dots, t_N)$ contains
a term of the form
\begin{align}
\phi_{i\,\text{(inst)}}^{i_1\dots i_N}(t,t_1,\dots,
t_N)\equiv\mathcal{I}_{i}^{i_1\dots
i_N}(t)\prod_{\ell=1}^N\delta(t-t_\ell).\label{ginst}
\end{align}

One of the main results of this work is the following formula that gives
$\mathcal{I}_{i}^{i_1\dots i_N}(t)$ as the expectation value of
$\hat{H}_{i_1\dots i_N}(t)$ in Eq.~\eqref{defHN2}:
\begin{align}
\mathcal{I}_{i}^{i_1\dots i_N}(t)=\lim_{t'\rightarrow
t-0}\sigma_{i}^{i_1\dots i_N}(t,t',\dots,
t')=\frac{1}{V}\text{tr}\big(\hat{H}_{ii_1\dots
i_N}(t)\hat{\rho}_0(t)\big).\label{main1}
\end{align}
See Sec.~\ref{fsumderivation} for the derivation. This relation can be interpreted as the generalized $f$-sum rule for
non-equilibrium states as we discuss in Sec.~\ref{subsecsumrule}.

\subsection{Fourier transformation for stationary states}
\label{subsecsumrule}
If the unperturbed Hamiltonian $\hat{H}_0$ lacks any time-dependence and
if the initial density matrix $\hat{\rho}_0$ commutes with $\hat{H}_0$,
the system becomes time-translation invariant.  In such a case,
$\hat{\rho}_0$ can be written as
\begin{align}
\hat{\rho}_0=\sum_n\rho_n|n\rangle\langle n|,\label{defrho0}
\end{align}
where $|n\rangle$ is the $N$-th eigenstate of $\hat{H}_0$ with the
energy eigenvalue $\mathcal{E}_n$.  In the stationary state, both
$\sigma_{i}^{i_1\dots i_N}(t,t_1,\dots, t_N)$ and $\phi_{i}^{i_1\dots
i_N}(t,t_1,\dots, t_N)$ depend only on the difference $t-t_\ell$. We
write
\begin{align}
&\sigma_{i}^{i_1\dots i_N}(t,t_1,\dots, t_N)=\sigma_{i}^{i_1\dots
i_N}(t-t_1,\dots, t-t_N),\\ &\phi_{i}^{i_1\dots i_N}(t,t_1,\dots,
t_N)=\phi_{i}^{i_1\dots i_N}(t-t_1,\dots, t-t_N),
\end{align}
for which Eq.~\eqref{gtosigma} becomes
\begin{align}
\sigma_{i}^{i_1\dots i_N}(t_1,\dots, t_N)=\int_{-\epsilon}^{t_1}dt_1'
\dots \int_{-\epsilon}^{t_N}dt_N' \phi_{i}^{i_1\dots i_N}(t_1',\dots,
t_N').\label{gtosigma2}
\end{align}

In the presence of the time-translation symmetry, it is customary to
work in the Fourier space. The Fourier transformation is defined by
\begin{align}
&\sigma_i^{i_1\dots
i_N}(\omega_1,\dots,\omega_N)=\int_{-\epsilon}^{\infty}dt_1\dots
\int_{-\epsilon}^{\infty}dt_N\sigma_i^{i_1\dots
i_N}(t_1,\dots,t_N)\prod_{\ell=1}^Ne^{(i\omega_\ell-\eta)t_\ell},\label{deffourier1}\\
&\phi_i^{i_1\dots
i_N}(\omega_1,\dots,\omega_N)=\int_{-\epsilon}^{\infty}dt_1\dots
\int_{-\epsilon}^{\infty}dt_N\phi_i^{i_1\dots
i_N}(t_1,\dots,t_N)\prod_{\ell=1}^Ne^{(i\omega_\ell-\eta)t_\ell},\label{deffourier2}
\end{align}
where $\eta>0$ is an infinitesimal parameter ensuring the convergence of
the integrand in the $t_\ell\to\infty$ limit. The Fourier component
$\sigma_i^{i_1\dots i_N}(\omega_1,\dots,\omega_N)$ is called the $N$-th
order optical conductivity.  In the Fourier space, Eq.~\eqref{gtosigma2}
reduces to
\begin{align}
\sigma_i^{i_1\dots i_N}(\omega_1,\dots,\omega_N)=\phi_i^{i_1\dots
i_N}(\omega_1,\dots,\omega_N)\prod_{\ell=1}^N\frac{i}{\omega_\ell+i\eta}.\label{gtosigma3}
\end{align}

By the inverse Fourier transformation, the frequency integral of
$\sigma_{i}^{i_1\dots i_N}(\omega_1,\dots, \omega_N)$ can be related to
the instantaneous conductivity in Eq.~\eqref{main1} that is
time-independent in stationary states:
\begin{align}
&\int_{-\infty}^{\infty}\frac{d\omega_1}{2\pi}\dots
\int_{-\infty}^{\infty}\frac{d\omega_N}{2\pi}\sigma_i^{i_1\dots
i_N}(\omega_1,\dots,\omega_N)\notag\\
&=\frac{1}{2^N}\mathcal{I}_{i}^{i_1\dots
i_N}=\frac{1}{2^N}\lim_{t'\rightarrow +0}\sigma_{i}^{i_1\dots
i_N}(t',\dots, t')=\frac{1}{2^NV}\sum_n\rho_n\big\langle
n\big|\hat{H}_{ii_1\dots i_N}\big|n\big\rangle.~\label{fsumrule}
\end{align}
The factor $2^{-N}$ originates from the discontinuity of
$\sigma_{i}^{i_1\dots i_N}(t_1,\dots, t_N)$ around $t_\ell=0$.  This is
the generalized $f$-sum rule for non-linear conductivity proposed
recently~\cite{2003.10390}.

\subsection{Drude weight and Kohn formula}
The Drude weight, usually discussed for the linear optical conductivity
as in Eq.~\eqref{standardform}, can be naturally extended to nonlinear
responses.  The $N$-th order Drude weight ($N\geq1$) is the coefficient
of the term proportional to $\prod_{\ell=1}^Ni/(\omega_\ell+i\eta)$ in
the $N$-th order optical conductivity~\cite{2003.10390}:
\begin{align}
\sigma_{i\,\text{(Drude)}}^{i_1\dots
i_N}(\omega_1,\dots,\omega_N)=\mathcal{D}_{i}^{i_1\dots
i_N}\prod_{\ell=1}^N\frac{i}{\omega_\ell+i\eta}.\label{defDrude}
\end{align}
This is the most singular part of $\sigma_{i}^{i_1\dots i_N}(\omega_1,\dots,\omega_N)$ around $\omega_1=\dots=\omega_N=0$ proportional to $\prod_{i=1}^N\delta(\omega_\ell)$. One may think of $\sigma_{i\,\text{(Drude)}}^{i_1\dots
i_N}(\omega_1,\dots,\omega_N)$ as the long-time average of $\sigma_{i}^{i_1\dots
i_N}(t_1,\dots,t_N)$, in sharp contrast to the instantaneous value $\mathcal{I}_{i}^{i_1\dots i_N}(t)$ defined in Eq.~\eqref{sigmainst}.

Comparing with Eq.~\eqref{gtosigma3}, we find
\begin{equation}
\mathcal{D}_{i}^{i_1\dots i_N}= \phi_i^{i_1\dots
i_N}(\omega_1=0,\dots,\omega_N=0)=\int_{-\epsilon}^{\infty}dt_1\dots
\int_{-\epsilon}^{\infty}dt_N\phi_i^{i_1\dots
i_N}(t_1,\dots,t_N)e^{-\eta \sum_{\ell=1}^Nt_\ell}.
\end{equation}

The generalized Kohn formula proposed in Ref.~\cite{2003.10390} reads
\begin{align}
&\mathcal{D}_i^{i_1\dots i_N}=\frac{1}{V}\sum_n
\rho_n\frac{\partial^{N+1}\mathcal{E}_n(\vec{A})}{\partial A_{i}\partial
A_{i_1}\dots \partial A_{i_N}}\Big|_{\vec{A}=\vec{0}}.\label{main2}
\end{align}
Here, $\mathcal{E}_n(\vec{A})$ is the energy eigenvalue of the
(instantaneous) eigenstate $|n(\vec{A})\rangle$ of $\hat{H}(\vec{A})$.
The weight $\rho_n$ appearing in $\hat{\rho}_0$ in Eq.~\eqref{defrho0}
\emph{is kept independent of} $\vec{A}$.  The energy eigenvalue $\mathcal{E}_n(\vec{A})$ and
the eigenstate $|n(\vec{A})\rangle$ are assumed to be analytic around $\vec{A}=0$,
satisfying $\mathcal{E}_n(\vec{0})=\mathcal{E}_n$ and
$|n(\vec{0})\rangle=|n\rangle$.  The case of $N=1$ for Gibbs states
(i.e., $\rho_n\propto e^{-\beta \mathcal{E}_n}$) reduces to the
finite-temperature extension of the original Kohn formula discussed in
Ref.~\cite{PhysRevLett.74.972}, and Eq.~\eqref{main2} is its
generalization to the $N$-th order optical conductivity ($N\geq 1$) of
general stationary states. In Sec.~\ref{dwderivation}, we give a proof of
Eq.~\eqref{main2} via the non-degenerate perturbation theory for $N=2$ and $3$.

As a by-product of this derivation, we obtain the following alternative
formulas of the Drude weight up to the third-order response [see Eqs.~\eqref{D1}, \eqref{D2}, and \eqref{D3}]:
\begin{align}
\mathcal{D}_i^{i_1}&=\frac{1}{V}\sum_n\rho_n\Big(\langle n|
\hat{H}_{ii_1}|n\rangle-\langle
n|\hat{H}_i\frac{\hat{Q}_n}{\hat{H}_0-\mathcal{E}_{n}}\hat{H}_{i_1}|n\rangle+\text{c.c.}\Big),\label{druderl1}\\
\mathcal{D}_i^{i_1i_2}&=\mathcal{S}_{ii_1i_2}\frac{1}{V}\sum_n\rho_n\Big[\langle
n|\hat{H}_{ii_1i_2}|n\rangle-3\langle
n|\hat{H}_i\frac{\hat{Q}_n}{\hat{H}_0-\mathcal{E}_{n}}\hat{H}_{i_1i_2}|n\rangle+\text{c.c.}+6\langle
n|\hat{H}_{i_1}\frac{\hat{Q}_n}{\hat{H}_0-\mathcal{E}_{n}}\delta_n\hat{H}_{i}\frac{\hat{Q}_n}{\hat{H}_0-\mathcal{E}_{n}}\hat{H}_{i_2}|n\rangle\Big],\label{druderl2}\\
\mathcal{D}_{i}^{i_1i_2i_3}&=\mathcal{S}_{ii_1i_2i_3}\frac{1}{V}\sum_n\rho_n\Big[\langle
n|\hat{H}_{i i_1 i_2 i_3}|n\rangle-6\langle
n|\hat{H}_{ii_1}\frac{\hat{Q}_n}{\hat{H}_0-\mathcal{E}_n}\hat{H}_{i_2i_3}|n\rangle-4\langle
n|\hat{H}_{i}\frac{\hat{Q}_n}{\hat{H}_0-\mathcal{E}_n}\hat{H}_{i_1i_2i_3}
|n\rangle+\text{c.c.}\Big]\notag\\
&\quad+12\mathcal{S}_{ii_1i_2i_3}\frac{1}{V}\sum_n\rho_n\Big[ \langle
n|\hat{H}_{i_3}\frac{\hat{Q}_n}{\hat{H}_0-\mathcal{E}_{n}}\delta_n\hat{H}_{i_1i_2}\frac{\hat{Q}_n}{\hat{H}_0-\mathcal{E}_{n}}\hat{H}_i|n\rangle
+\langle
n|\hat{H}_{i_1i_2}\frac{\hat{Q}_n}{\hat{H}_0-\mathcal{E}_{n}}\delta_n\hat{H}_{i}\frac{\hat{Q}_n}{\hat{H}_0-\mathcal{E}_{n}}\hat{H}_{i_3}|n\rangle+\text{c.c.}\Big]\notag\\
&\quad-24\mathcal{S}_{ii_1i_2i_3}\frac{1}{V}\sum_n\rho_n\langle
n|\hat{H}_i\frac{\hat{Q}_n}{\hat{H}_0-\mathcal{E}_{n}}\delta_n\hat{H}_{i_1}\frac{\hat{Q}_n}{\hat{H}_0-\mathcal{E}_{n}}\delta_n\hat{H}_{i_2}\frac{\hat{Q}_n}{\hat{H}_0-\mathcal{E}_{n}}\hat{H}_{i_3}|n\rangle\notag\\
&\quad+24\mathcal{S}_{ii_1i_2i_3}\frac{1}{V}\sum_n\rho_n\langle
n|\hat{H}_i\frac{\hat{Q}_n}{(\hat{H}_0-\mathcal{E}_{n})^2}\hat{H}_{i_1}|n\rangle\langle
n|\hat{H}_{i_2}\frac{\hat{Q}_n}{\hat{H}_0-\mathcal{E}_{n}}\hat{H}_{i_3}|n\rangle.\label{druderl3}
\end{align}
Here, $\hat{Q}_n\equiv1-|n\rangle\langle n|$ is the projector onto the
compliment of the space spanned by $|n\rangle$, $\text{c.c.}$ represents
the complex conjugation of the term right in front,
$\mathcal{S}_{ii_1\dots i_N}$ is the symmetrizing operation among $i$,
$i_1$, $\dots$, and $i_N$, and $\delta_n\hat{H}_i$ is a short-hand
notation for $\hat{H}_i-\langle n|\hat{H}_i|n\rangle$.  The expression
~\eqref{druderl1} for the linear Drude weight has been used in the
literature~\cite{Kohn1964,PhysRevLett.74.972,Resta_2018}, and
Eqs.~\eqref{druderl2} and \eqref{druderl3} are its generalizations to
the second- and the third order Drude weight.  The advantage of these
expressions is that the gauge field $\vec{A}$ is set to be $\vec{0}$
when diagonalizing the Hamiltonian $\hat{H}_0$ to find $|n\rangle$ and
$\mathcal{E}_n$.  As a price to pay, one needs to know all
excited states to compute the correlation functions. This is in contrast
to the Kohn formula \eqref{main2} at zero temperature, which uses only
the ground state energy as a function of $\vec{A}$.

\section{Kubo theory}
In this section, we derive the concrete expressions of $\phi_i^{i_1\dots
i_N}(t,t_1,\dots,t_N)$ for $N=1$, $2$ and $3$ by following Kubo's work
on the response theory~\cite{Kubo}.  This formulation naturally
leads us to a proof of Eq.~\eqref{main1} on the instantaneous
conductivity for a general $N\geq1$.

Let us begin by expanding the perturbed Hamiltonian
$\hat{H}(t,\vec{A}(t))$ as
\begin{align}
\hat{H}(t,\vec{A}(t))=\sum_{N=0}^\infty\hat{H}_{N}(t),\label{expandH}
\end{align}
where $\hat{H}_{N}(t)$ is the correction of all $N$-th order terms,
\begin{align}
&\hat{H}_{N}(t)\equiv \frac{1}{N!}\sum_{i_1,\dots,i_N}\hat{H}_{i_1\dots
i_N}(t)\prod_{\ell=1}^NA_{i_\ell}(t)\label{defHN}.
\end{align}
The perturbed density matrix $\hat{\rho}(t)$ can be accordingly written
as
\begin{align}
&\hat{\rho}(t)=\sum_{N=0}^\infty\hat{\rho}_{N}(t).\label{expandrho}
\end{align}
Plugging Eqs.~\eqref{expandH} and \eqref{expandrho} into the von Neumann
equation
$i\partial_t\hat{\rho}(t)=[\hat{H}(t,\vec{A}(t)),\hat{\rho}(t)]$, we
find, order by order, that
\begin{align}
\partial_t\hat{\rho}_{N}(t)&=\sum_{M=0}^N(-i)[\hat{H}_{N-M}(t),\hat{\rho}_M(t)].
\end{align}
We can solve this equation for $\hat{\rho}_N(t)$ by switching to the
``interacting picture"
\begin{align}
\hat{\tilde{O}}(t)\equiv
\hat{S}_0^\dagger(t)\hat{O}(t)\hat{S}_0(t),\label{intpic}
\end{align}
where $\hat{S}_0(t)$ was defined in Eq.~\eqref{defS0}. Assuming
$\hat{\rho}_N(0)=0$ for $N\geq1$, we find
\begin{align}
\hat{\tilde{\rho}}_N(t)&=\sum_{M=0}^{N-1}\int_0^{t}
dt'(-i)[\hat{\tilde{H}}_{N-M}(t'),\hat{\tilde{\rho}}_M(t')]\quad\text{($N\geq
1$)}.\label{rhon}
\end{align}
Since the right-hand side contains only $\hat{\tilde{\rho}}_M(t')$ with
$0\leq M\leq N-1$, one can find $\hat{\tilde{\rho}}_N(t)$
inductively. For example, we have
$\hat{\tilde{\rho}}_0(t)=\hat{\rho}_0$,
\begin{align}
\hat{\tilde{\rho}}_1(t)&=\int_0^{t}
dt'(-i)[\hat{\tilde{H}}_{1}(t'),\hat{\rho}_0],\label{rho1}
\end{align}
\begin{align}
\hat{\tilde{\rho}}_2(t)&=\int_0^{t}
dt'(-i)[\hat{\tilde{H}}_{2}(t'),\hat{\rho}_0]+\int_0^{t} dt'\int_0^{t'}
dt''(-i)^2[\hat{\tilde{H}}_{1}(t'),[\hat{\tilde{H}}_{1}(t''),\hat{\rho}_0]],\label{rho2}
\end{align}
and
\begin{align}
\hat{\tilde{\rho}}_3(t)&=\int_0^{t}
dt'(-i)[\hat{\tilde{H}}_{3}(t'),\hat{\rho}_0]+\int_0^{t} dt'\int_0^{t'}
dt''(-i)^2\big([\hat{\tilde{H}}_{2}(t'),[\hat{\tilde{H}}_{1}(t''),\hat{\rho}_0]]+[\hat{\tilde{H}}_{1}(t'),[\hat{\tilde{H}}_{2}(t''),\hat{\rho}_0]\big)\notag\\
&\quad+\int_0^{t} dt'\int_0^{t'} dt''\int_0^{t''}
dt'''(-i)^3[\hat{\tilde{H}}_{1}(t'),[\hat{\tilde{H}}_{1}(t''),[\hat{\tilde{H}}_{1}(t'''),\hat{\rho}_0]]].\label{rho3}
\end{align}
Expressions for higher-order terms can be derived, at least formally, in
the same way. However, when expressed in terms of $\hat{H}_{M}$ ($0\leq
M\leq N$) and $\hat{\rho}_0$, the number of terms in
$\hat{\tilde{\rho}}_N(t)$ ($N\geq1$) is $2^{N-1}$, and it is practically
not easy to keep track of them altogether. In what follows, we will
mainly discuss corrections only up to the third order ($N=3$).

The current density operator in Eq.~\eqref{defj} depends on $\vec{A}(t)$
explicitly, which can also be written as
\begin{align}
&\hat{j}_i(t,\vec{A}(t))=\sum_{N=0}^\infty\hat{j}_{iN}(t),\\
&\hat{j}_{iN}(t)\equiv\frac{1}{V}\sum_{N=0}^\infty\frac{1}{N!}\sum_{i_1,\dots,i_N}\hat{H}_{ii_1\dots
i_{N}}(t)\prod_{\ell=1}^{N}A_{i_\ell}(t).\label{defjN}
\end{align}
The $N$-th order correction to the current expectation value is the sum
of $N+1$ contributions:
\begin{align}
j_{i}^{(N)}(t)=\sum_{M=0}^N\text{tr}\big(\hat{\tilde{j}}_{i
N-M}(t)\hat{\tilde{\rho}}_M(t)\big),\label{jiNM}
\end{align}
which contains $2^N$ terms in total. Using
Eq.~\eqref{rho1}--\eqref{rho3}, we get
\begin{align}
j_{i}^{(1)}(t)&=\langle\hat{\tilde{j}}_{i 1}(t)\rangle_0+\int_0^{t}
dt'(-i)\langle[\hat{\tilde{j}}_{i
0}(t),\hat{\tilde{H}}_{1}(t')]\rangle_0,\label{ji1}
\end{align}
\begin{align}
j_{i}^{(2)}(t)&=\langle\hat{\tilde{j}}_{i 2}(t)\rangle_0+\int_0^{t}
dt'(-i)\langle[\hat{\tilde{j}}_{i
1}(t),\hat{\tilde{H}}_{1}(t')]\rangle_0\notag\\ &\quad+\int_0^{t}
dt'(-i)\langle[\hat{\tilde{j}}_{i
0}(t),\hat{\tilde{H}}_{2}(t')]\rangle_0+\int_0^{t} dt'\int_0^{t'}
dt''(-i)^2\langle[[\hat{\tilde{j}}_{i
0}(t),\hat{\tilde{H}}_{1}(t')],\hat{\tilde{H}}_{1}(t'')]\rangle_0,\label{ji2}
\end{align}
and
\begin{align}
j_{i}^{(3)}(t)&=\langle\hat{\tilde{j}}_{i 3}(t)\rangle_0+\int_0^{t}
dt'(-i)\langle[\hat{\tilde{j}}_{i
2}(t),\hat{\tilde{H}}_{1}(t')]\rangle_0\notag\\ &\quad+\int_0^{t}
dt'(-i)\langle[\hat{\tilde{j}}_{i
1}(t),\hat{\tilde{H}}_{2}(t')]\rangle_0+\int_0^{t} dt'\int_0^{t'}
dt''(-i)^2\langle[[\hat{\tilde{j}}_{i
1}(t),\hat{\tilde{H}}_{1}(t')],\hat{\tilde{H}}_{1}(t'')]\rangle_0\notag\\
&\quad+\int_0^{t} dt'(-i)\langle[\hat{\tilde{j}}_{i
0}(t),\hat{\tilde{H}}_{3}(t')]\rangle_0+\int_0^{t} dt'\int_0^{t'}
dt''(-i)^2\big(\langle[[\hat{\tilde{j}}_{i
0}(t),\hat{\tilde{H}}_{2}(t')],\hat{\tilde{H}}_{1}(t'')]\rangle_0+\langle[[\hat{\tilde{j}}_{i
0}(t),\hat{\tilde{H}}_{1}(t')],\hat{\tilde{H}}_{2}(t'')]\rangle_0\big)\notag\\
&\quad+\int_0^{t} dt'\int_0^{t'} dt''\int_0^{t''}
dt'''(-i)^3\langle[[[\hat{\tilde{j}}_{i
0}(t),\hat{\tilde{H}}_{1}(t')],\hat{\tilde{H}}_{1}(t'')],\hat{\tilde{H}}_{1}(t''')]\rangle_0.\label{ji3}
\end{align}
Here and hereafter, we write
\begin{align}
\langle\hat{O}\rangle_0\equiv
\text{tr}\big(\hat{O}\hat{\rho}_0\big)\label{expectation0}
\end{align}
for any operator $\hat{O}$.  Plugging Eqs.~\eqref{defHN} and
\eqref{defjN} into these results, we can read off $\phi_{i}^{i_1\dots
i_N}(t,t_1,\dots,t_N)$:
\begin{align}
V\phi_{i}^{i_1}(t,t_1)&=\delta(t-t_1)\langle\hat{\tilde{H}}_{i
i_1}(t)\rangle_0\notag\\
&\quad+\theta(t-t_1)(-i)\langle[\hat{\tilde{H}}_{i}(t),\hat{\tilde{H}}_{i_1}(t_1)]\rangle_0,\label{gtt1}
\end{align}
\begin{align}
V\phi_{i}^{i_1i_2}(t,t_1,t_2)&=\delta(t-t_1)\delta(t-t_2)\langle\hat{\tilde{H}}_{ii_1i_2}(t)\rangle_0\notag\\
&\quad+2\mathcal{S}_{i_1i_2}\delta(t-t_1)\theta(t-t_2)(-i)\langle[\hat{\tilde{H}}_{i
i_1}(t),\hat{\tilde{H}}_{i_2}(t_2)]\rangle_0\notag\\
&\quad+\delta(t_1-t_2)\theta(t-t_1)(-i)\langle[\hat{\tilde{H}}_{i}(t),\hat{\tilde{H}}_{i_1i_2}(t_1)]\rangle_0\notag\\
&\quad+2\mathcal{S}_{i_1i_2}\theta(t-t_1)\theta(t_1-t_2)(-i)^2\langle[[\hat{\tilde{H}}_{i}(t),\hat{\tilde{H}}_{i_1}(t_1)],\hat{\tilde{H}}_{i_2}(t_2)]\rangle_0,\label{gtt2}
\end{align}
and
\begin{align}
V\phi_{i}^{i_1i_2i_3}(t,t_1,t_2,t_3)
&=\delta(t-t_1)\delta(t-t_2)\delta(t-t_3)\langle\hat{\tilde{H}}_{i i_1
i_2 i_3}(t)\rangle_0\notag\\
&\quad+3\mathcal{S}_{i_1i_2i_3}\delta(t-t_1)\delta(t-t_2)\theta(t-t_3)(-i)\langle[\hat{\tilde{H}}_{ii_1i_2}(t),\hat{\tilde{H}}_{i_3}(t_3)]\rangle_0\notag\\
&\quad+3\mathcal{S}_{i_1i_2i_3}\delta(t-t_1)\delta(t_2-t_3)\theta(t-t_2)(-i)\langle[\hat{\tilde{H}}_{i
i_1}(t),\hat{\tilde{H}}_{i_2i_3}(t_2)]\rangle_0\notag\\
&\quad+6\mathcal{S}_{i_1i_2i_3}\delta(t-t_1)\theta(t-t_2)\theta(t_2-t_3)(-i)^2\langle[[\hat{\tilde{H}}_{i
i_1}(t),\hat{\tilde{H}}_{i_2}(t_2)],\hat{\tilde{H}}_{i_3}(t_3)]\rangle_0\notag\\
&\quad+\delta(t_1-t_2)\delta(t_2-t_3)\theta(t-t_1)(-i)\langle[\hat{\tilde{H}}_{i}(t),\hat{\tilde{H}}_{i_1i_2i_3}(t_1)]\rangle_0\notag\\
&\quad+3\mathcal{S}_{i_1i_2i_3}\delta(t_1-t_2)\theta(t-t_1)\theta(t_1-t_3)(-i)^2\langle[[\hat{\tilde{H}}_{i}(t),\hat{\tilde{H}}_{i_1i_2}(t_1)],\hat{\tilde{H}}_{i_3}(t_3)]\rangle_0\notag\\
&\quad+3\mathcal{S}_{i_1i_2i_3}\delta(t_2-t_3)\theta(t-t_1)\theta(t_1-t_2)(-i)^2\langle[[\hat{\tilde{H}}_{i}(t),\hat{\tilde{H}}_{i_1}(t_1)],\hat{\tilde{H}}_{i_2i_3}(t_2)]\rangle_0\notag\\
&\quad+6\mathcal{S}_{i_1i_2i_3}\theta(t-t_1)\theta(t_1-t_2)\theta(t_2-t_3)(-i)^3\langle[[[\hat{\tilde{H}}_{i}(t),\hat{\tilde{H}}_{i_1}(t_1)],\hat{\tilde{H}}_{i_2}(t_2)],\hat{\tilde{H}}_{i_3}(t_3)]\rangle_0.\label{gtt3}
\end{align}
Here, $\theta(t)$ is the Heaviside step function and
$\mathcal{S}_{i_1\dots i_N}$ refers to the symmetrizing operation
averaging over all $N!$ permutations of $(i_\ell,t_\ell)$'s. For
example,
\begin{align}
\mathcal{S}_{i_1i_2}f_{i_1i_2}(t_1,t_2)&=\frac{1}{2}\big[f_{i_1i_2}(t_1,t_2)+f_{i_2i_1}(t_2,t_1)\big],\label{symmetrizing1}\\
\mathcal{S}_{i_1i_2i_3}f_{i_1i_2i_3}(t_1,t_2,t_3)&=\frac{1}{6}\big[f_{i_1i_2i_3}(t_1,t_2,t_3)+f_{i_2i_3i_1}(t_2,t_3,t_1)+f_{i_3i_1i_2}(t_3,t_1,t_2)\notag\\
&\quad\quad\quad+f_{i_3i_2i_1}(t_3,t_2,t_1)+f_{i_1i_3i_2}(t_1,t_3,t_2)+f_{i_2i_1i_3}(t_2,t_1,t_3)\big].\label{symmetrizing2}
\end{align}
One can obtain the expression of the corresponding $\sigma_{i}^{i_1\dots
i_N}(t,t_1,\dots,t_N)$ by using Eq.~\eqref{gtosigma}.

Let us switch back to the consideration for a general $N\geq1$. In
general, $j_{i}^{(N)}(t)$ ($N\geq0$) contains $\langle\hat{\tilde{j}}_{i
N}(t)\rangle_0=\langle\hat{\tilde{H}}_{i i_1 \dots i_N}(t)\rangle_0/V$
that originates from the $M=0$ contribution in Eq.~\eqref{jiNM}. This
term results in the instantaneous response
\begin{align}
\phi_{i\,\text{(inst)}}^{i_1\dots
i_N}(t,t_1,\dots,t_N)=\frac{1}{V}\langle\hat{\tilde{H}}_{i i_1 \dots
i_N}(t)\rangle_0\prod_{\ell=1}^N\delta(t-t_\ell)=\frac{1}{V}\text{tr}\big(\hat{H}_{i
i_1 \dots
i_N}(t)\hat{\rho}_0(t)\big)\prod_{\ell=1}^N\delta(t-t_\ell).\label{ginst2}
\end{align}
All other contributions to $j_{i}^{(N)}(t)$ from $M\geq 1$ in
Eq.~\eqref{jiNM} are retarded effect involving at least one temporal
integration that can be traced back to Eq.~\eqref{rhon}.  Comparing
Eq.~\eqref{ginst2} with Eq.~\eqref{ginst}, we obtain our result in
Eq.~\eqref{main1}.

\section{Direct proof of the generalized Kohn formula}
\label{Kohn}

In this section we perform the Fourier transformation to obtain the
explicit form of $\phi_i^{i_1\dots i_N}(\omega_1,\dots,\omega_N)$ for
$N=1$, $2$ and $3$. Then we use them to give a perturbative proof of the
Kohn formula for the second- and third-order optical conductivity.

\subsection{Optical conductivity and $f$-sum rules}
\label{fsumderivation}
In the presence of the time-translation symmetry,
Eqs.~\eqref{gtt1} and \eqref{gtt2} reduce to
\begin{align}
V\phi_i^{i_1}(t_1)&= \delta(t_1)\langle\hat{H}_{ii_1}\rangle_0\notag\\
&\quad+\theta(t_1)(-i)\langle[\hat{H}_{i},e^{-i\hat{H}_0t_1}\hat{H}_{i_1}e^{i\hat{H}_0t_1}]\rangle_0\label{gt1}
\end{align}
and
\begin{align}
V\phi_{i}^{i_1i_2}(t_1,t_2)&=\delta(t_1)\delta(t_2)\langle\hat{H}_{ii_1i_2}\rangle_0\notag\\
&\quad+2\mathcal{S}_{i_1i_2}\delta(t_1)\theta(t_2)(-i)\langle[\hat{H}_{ii_1},e^{-i\hat{H}_0t_2}\hat{H}_{i_2}e^{i\hat{H}_0t_2}]\rangle_0\notag\\
&\quad+\delta(t_2-t_1)\theta(t_1)(-i)\langle[\hat{H}_i,e^{-i\hat{H}_0t_1}\hat{H}_{i_1i_2}e^{i\hat{H}_0t_1}]\rangle_0\notag\\
&\quad+2\mathcal{S}_{i_1i_2}\theta(t_1)\theta(t_2-t_1)(-i)^2\langle[[\hat{H}_i,e^{-i\hat{H}_0t_1}\hat{H}_{i_1}e^{i\hat{H}_0t_1}],e^{-i\hat{H}_0t_2}\hat{H}_{i_2}e^{i\hat{H}_0t_2}]\rangle_0.\label{gt2}
\end{align}
We perform the Fourier transformation \eqref{deffourier2} assuming the
form of the density matrix in Eq.~\eqref{defrho0}. We find
\begin{align}
V\phi_i^{i_1}(\omega_1)&=\langle \hat{H}_{ii_1}\rangle_0\notag\\
&\quad+\sum_n\rho_n\langle
n|\Big(\delta_n\hat{H}_i\frac{\hat{Q}_n}{\omega_1-\hat{H}_0+\mathcal{E}_{n}+i\eta}\delta_n\hat{H}_{i_1}-\delta_n\hat{H}_{i_1}\frac{\hat{Q}_n}{\omega_1+\hat{H}_0-\mathcal{E}_{n}+i\eta}\delta_n\hat{H}_{i}\Big)|n\rangle
\end{align}
and
\begin{align}
V\phi_{i}^{i_1i_2}(\omega_1,\omega_2) &=\langle
\hat{H}_{ii_1i_2}\rangle_0\notag\\
&\quad+2\mathcal{S}_{i_1i_2}\sum_n\rho_n\langle
n|\Big(\delta_n\hat{H}_{ii_1}\frac{\hat{Q}_n}{\omega_2-\hat{H}_0+\mathcal{E}_{n}+i\eta}\delta_n\hat{H}_{i_2}-\delta_n\hat{H}_{i_2}\frac{\hat{Q}_n}{\omega_2+\hat{H}_0-\mathcal{E}_{n}+i\eta}\delta_n\hat{H}_{ii_1}\Big)|n\rangle\notag\\
&\quad+\sum_n\rho_n\langle
n|\Big(\delta_n\hat{H}_i\frac{\hat{Q}_n}{\omega_1+\omega_2-\hat{H}_0+\mathcal{E}_{n}+2i\eta}\delta_n\hat{H}_{i_1i_2}-\delta_n\hat{H}_{i_1i_2}\frac{\hat{Q}_n}{\omega_1+\omega_2+\hat{H}_0-\mathcal{E}_{n}+2i\eta}\delta_n\hat{H}_{i}\Big)|n\rangle\notag\\
&\quad+2\mathcal{S}_{i_1i_2}\sum_n\rho_n\Big[\langle
n|\delta_n\hat{H}_i\frac{\hat{Q}_n}{\omega_1+\omega_2-\hat{H}_0+\mathcal{E}_{n}+2i\eta}\delta_n\hat{H}_{i_1}\frac{\hat{Q}_n}{\omega_2-\hat{H}_0+\mathcal{E}_{n}+i\eta}\delta_n\hat{H}_{i_2}|n\rangle\notag\\
&\quad\quad\quad\quad\quad\quad\quad\quad-\langle
n|\delta_n\hat{H}_{i_1}\frac{\hat{Q}_n}{\omega_1+\hat{H}_0-\mathcal{E}_{n}+i\eta}\delta_n\hat{H}_i\frac{\hat{Q}_n}{\omega_2-\hat{H}_0+\mathcal{E}_{n}+i\eta}\delta_n\hat{H}_{i_2}|n\rangle\notag\\
&\quad\quad\quad\quad\quad\quad\quad\quad+\langle
n|\delta_n\hat{H}_{i_2}\frac{\hat{Q}_n}{\omega_2+\hat{H}_0-\mathcal{E}_{n}+i\eta}\delta_n\hat{H}_{i_1}\frac{\hat{Q}_n}{\omega_1+\omega_2+\hat{H}_0-\mathcal{E}_{n}+2i\eta}\delta_n\hat{H}_i|n\rangle\Big].
\end{align}
Here,$\delta_n\hat{H}_i\equiv\hat{H}_i-\langle n|\hat{H}_i|n\rangle$,
$\hat{Q}_n\equiv1-|n\rangle\langle n|$ is the projector onto the
compliment of the space spanned by $|n\rangle$, and
$\mathcal{S}_{i_1\dots i_N}$ is the symmetrization among
$(i_\ell,\omega_\ell)$'s.  The
corresponding optical conductivity is then given by
Eq.~\eqref{gtosigma3}.
The expression of
$\phi_{i}^{i_1i_2i_3}(t_1,t_2,t_3)$ and $\phi_{i}^{i_1i_2i_3}(\omega_1,\omega_2,\omega_3)$ are too long to be
presented here and are included in Appendix~\ref{phithird}.

More generally, the instantaneous contribution in Eq.~\eqref{ginst2}
gives rise to
\begin{align}
&\sigma_{i\,\text{(inst)}}^{i_1\dots
i_N}(\omega_1,\dots,\omega_N)=\frac{1}{V}\langle \hat{H}_{ii_1\dots
i_N}\rangle_0\prod_{\ell=1}^N\frac{i}{\omega_\ell+i\eta}\label{instomega}
\end{align}
in the $N$-th other optical conductivity.  All other terms in
$\sigma_i^{i_1\dots i_N}(\omega_1,\dots,\omega_N)$ are suppressed by
$|\omega_\ell|^{-2}$ for large $|\omega_\ell|$ for at least one
$1\leq\ell\leq N$ as a result of the temporal integration in
Eq.~\eqref{rhon}.  This observation immediately implies the $f$-sum rule
\eqref{fsumrule}. To see this, let us perform the frequency integration
in Eq.~\eqref{fsumrule} using techniques of complex analysis.  Except
for the instantaneous term in Eq.~\eqref{instomega}, all other terms in
$\sigma_i^{i_1\dots i_N}(\omega_1,\dots,\omega_N)$ do not contribute to
this integral. This is because, if a term is suppressed by
$|\omega_\ell|^{-2}$ for large $|\omega_\ell|$, one can form a closed
integration path by adding a large half circle in the upper half of the
complex $\omega_\ell$-plane and apply the Cauchy's integral theorem.
All poles of $\sigma_i^{i_1\dots i_N}(\omega_1,\dots,\omega_N)$ are
located in the lower-half of the complex plane and thus the integral
vanishes.  Therefore, taking into account only the contribution from the
instantaneous term \eqref{instomega} using the formula
\begin{align}
\int_{-\infty}^{\infty}\frac{d\omega}{2\pi
i}\frac{i}{\omega+i\eta}=\theta(0)=\frac{1}{2},
\end{align}
we reproduce the $f$-sum rule \eqref{fsumrule}.

\subsection{Drude weight}
\label{dwderivation}
Here we derive the Kohn formula based on the concrete expressions of
$\phi_i^{i_1\dots i_N}(\omega_1,\dots,\omega_N)$ obtained above. For the
brevity of the presentation, let us write
\begin{align}
|n_{i_1\dots
 i_N}\rangle\equiv\frac{\partial^N|n(\vec{A})\rangle}{\partial
 A_{i_1}\dots\partial A_{i_N}}\Big|_{\vec{A}=\vec{0}}.
\end{align}

The linear Drude weight is given by
$\mathcal{D}_i^{i_1}=\phi_i^{i_1}(\omega_1=0)$.  Let us make this into
the form of the Kohn formula following the discussions in
Refs.~\onlinecite{Kohn1964,PhysRevLett.74.972,Resta_2018}.  Using the
standard formulas
\begin{align}
&\langle n|n_{i_1}\rangle+\text{c.c.}=0,\label{fo1}\\
&\hat{Q}_n|n_{i_1}\rangle=-\frac{\hat{Q}_n}{\hat{H}_0-\mathcal{E}_n}\hat{H}_{i_1}|n\rangle\label{fo2}
\end{align}
in the first-order nondegenerate perturbation theory, we find,
\begin{align}
V\mathcal{D}_i^{i_1}&=V\phi_i^{i_1}(\omega_1=0)\notag\\
&=\sum_n\rho_n\Big(\langle n| \hat{H}_{ii_1}|n\rangle-\langle
n|\hat{H}_i\frac{\hat{Q}_n}{\hat{H}_0-\mathcal{E}_{n}}\hat{H}_{i_1}|n\rangle+\text{c.c.}\Big)\notag\\
&= \sum_n\rho_n\big(\langle n| \hat{H}_{ii_1}|n\rangle+\langle
n|\hat{H}_i\hat{Q}_n|n_{i_1}\rangle+\text{c.c.}\big)\notag\\ &=
\sum_n\rho_n\big(\langle n| \hat{H}_{ii_1}|n\rangle+\langle
n|\hat{H}_i|n_{i_1}\rangle+\text{c.c.}\big)\notag\\
&=\sum_n\rho_n\frac{\partial}{\partial A_{i_1}}\Big\langle
n(\vec{A})\Big|\frac{\partial\hat{H}(\vec{A})}{\partial
A_{i}}\Big|n(\vec{A})\Big\rangle\Big|_{\vec{A}=\vec{0}}\notag\\
&=\sum_n\rho_n\frac{\partial^2\mathcal{E}_n(\vec{A})}{\partial
A_i\partial A_{i_1}}\Big|_{\vec{A}=\vec{0}}.\label{D1}
\end{align}
Here, $\text{c.c.}$ represents the complex conjugation of the term right
in front. Thus Eq.~\eqref{main2} for $N=1$ is verified.  Strictly
speaking, this proof applies only to the cases where all energy levels
$\mathcal{E}_n$ with $\rho_n\neq0$ are non-degenerate.  It was argued in
Ref.~\cite{PhysRevLett.74.972} that the first-order degenerate
perturbation theory lifts the degeneracy and the same procedure should
work.

Let us do the same for the second-order Drude weight, given by
$\mathcal{D}_i^{i_1i_2}=\phi_i^{i_1i_2}(\omega_1=0,\omega_2=0)$.
Knowing the answer, we find it easier to go backward:
\begin{align}
&\sum_n\rho_n\frac{\partial^3\mathcal{E}_n(\vec{A})}{\partial
A_i\partial A_{i_1}\partial A_{i_2}}\Big|_{\vec{A}=\vec{0}}\notag\\
&=\sum_n\rho_n\frac{\partial^2}{\partial A_{i_1}\partial
A_{i_2}}\Big\langle
n(\vec{A})\Big|\frac{\partial\hat{H}(\vec{A})}{\partial
A_{i}}\Big|n(\vec{A})\Big\rangle\Big|_{\vec{A}=\vec{0}}\notag\\
&=\mathcal{S}_{i_1i_2}\sum_n\rho_n\big(\langle n|
\hat{H}_{ii_1i_2}|n\rangle+2\langle
n_{i_1}|\hat{H}_{i}|n_{i_2}\rangle\big)+\mathcal{S}_{i_1i_2}\sum_n\rho_n\big(2\langle
n|\hat{H}_{ii_1}|n_{i_2}\rangle+\langle
n|\hat{H}_i|n_{i_1i_2}\rangle\big)+\text{c.c.}\notag\\
&=\mathcal{S}_{i_1i_2}\sum_n\rho_n\big(\langle n|
\hat{H}_{ii_1i_2}|n\rangle+2\langle
n_{i_1}|\hat{Q}_n\delta_n\hat{H}_i\hat{Q}_n|n_{i_2}\rangle\big)+\mathcal{S}_{i_1i_2}\sum_n\rho_n\big(2\langle
n|\hat{H}_{ii_1}\hat{Q}_n|n_{i_2}\rangle+\langle
n|\hat{H}_i\hat{Q}_n|n_{i_1i_2}\rangle\big)+\text{c.c.}\notag\\
&\quad+2\mathcal{S}_{i_1i_2}\sum_n\rho_n\langle
n_{i_1}|\hat{Q}_n\hat{H}_i|n\rangle\langle
n|n_{i_2}\rangle+\text{c.c.}\notag\\
&=\mathcal{S}_{ii_1i_2}\sum_n\rho_n\Big[\langle
n|\hat{H}_{ii_1i_2}|n\rangle-3\langle
n|\hat{H}_i\frac{\hat{Q}_n}{\hat{H}_0-\mathcal{E}_{n}}\hat{H}_{i_1i_2}|n\rangle+\text{c.c.}+6\langle
n|\hat{H}_{i_1}\frac{\hat{Q}_n}{\hat{H}_0-\mathcal{E}_{n}}\delta_n\hat{H}_{i}\frac{\hat{Q}_n}{\hat{H}_0-\mathcal{E}_{n}}\hat{H}_{i_2}|n\rangle\Big]\notag\\
&=V\phi_i^{i_1i_2}(\omega_1=0,\omega_2=0)=V\mathcal{D}_i^{i_1i_2}.
\label{D2}
\end{align}
This reproduces Eq.~\eqref{main2} for $N=2$.  In the derivation, we used
the standard formulas in Eqs.~\eqref{fo1}, \eqref{fo2}, and
\begin{align}
&\left(\langle n|n_{i_1i_2}\rangle+\langle
n_{i_2}|n_{i_1}\rangle\right)+\text{c.c.}=0,\\
&\hat{Q}_n|n_{i_1i_2}\rangle=-\mathcal{S}_{i_1i_2}\left[\frac{\hat{Q}_n}{\hat{H}_0-\mathcal{E}_n}\hat{H}_{i_1i_2}|n\rangle+2\frac{\hat{Q}_n}{\hat{H}_0-\mathcal{E}_n}\delta_n\hat{H}_{i_1}|n_{i_2}\rangle\right].
\end{align}

Finally, for the third-order response, we have
\begin{align}
&\sum_n\rho_n\frac{\partial^4\mathcal{E}_n(\vec{A})}{\partial
A_i\partial A_{i_1}\partial A_{i_2}\partial
A_{i_3}}\Big|_{\vec{A}=\vec{0}}\notag\\
&=\sum_n\rho_n\frac{\partial^3}{\partial A_{i_1}\partial A_{i_2}\partial
A_{i_3}}\Big\langle
n(\vec{A})\Big|\frac{\partial\hat{H}(\vec{A})}{\partial
A_{i}}\Big|n(\vec{A})\Big\rangle\Big|_{\vec{A}=\vec{0}}\notag\\
&=\mathcal{S}_{i_1i_2i_3}\sum_n\rho_n\big(\langle n|
\hat{H}_{ii_1i_2i_3}|n\rangle+6\langle
n_{i_2}|\hat{H}_{ii_1}|n_{i_3}\rangle\big)\notag\\
&\quad+\mathcal{S}_{i_1i_2i_3}\sum_n\rho_n\big(3\langle
n|\hat{H}_{ii_1i_2}|n_{i_3}\rangle+3\langle
n|\hat{H}_{ii_1}|n_{i_2i_3}\rangle+3\langle
n_{i_1}|\hat{H}_{i}|n_{i_2i_3}\rangle+\langle
n|\hat{H}_{i}|n_{i_1i_2i_3}\rangle\big)+\text{c.c.}\notag\\
&=\mathcal{S}_{i_1i_2i_3}\sum_n\rho_n\big(\langle n|
\hat{H}_{ii_1i_2i_3}|n\rangle+6\langle
n_{i_2}|\hat{Q}_n\delta_n\hat{H}_{ii_1}\hat{Q}_n|n_{i_3}\rangle\big)\notag\\
&\quad+\mathcal{S}_{i_1i_2i_3}\sum_n\rho_n\big(3\langle
n|\hat{H}_{ii_1i_2}\hat{Q}_n|n_{i_3}\rangle+3\langle
n|\hat{H}_{ii_1}\hat{Q}_n|n_{i_2i_3}\rangle+3\langle
n_{i_1}|\hat{Q}_n\delta_n\hat{H}_{i}\hat{Q}_n|n_{i_2i_3}\rangle+\langle
n|\hat{H}_{i}\hat{Q}_n|n_{i_1i_2i_3}\rangle\big)+\text{c.c.}\notag\\
&\quad+\mathcal{S}_{i_1i_2i_3}\sum_n\rho_n\big(6\langle
n_{i_2}|n\rangle\langle
n|\hat{H}_{ii_1}\hat{Q}_n|n_{i_3}\rangle+3\langle
n_{i_1}|n\rangle\langle
n|\hat{H}_{i}\hat{Q}_n|n_{i_2i_3}\rangle+3\langle
n_{i_1}|\hat{Q}_n\hat{H}_{i}|n\rangle\langle
n|n_{i_2i_3}\rangle\big)+\text{c.c.}\notag\\
&=\mathcal{S}_{ii_1i_2i_3}\sum_n\rho_n\Big[\langle n|\hat{H}_{i i_1 i_2
i_3}|n\rangle-6\langle
n|\hat{H}_{ii_1}\frac{\hat{Q}_n}{\hat{H}_0-\mathcal{E}_n}\hat{H}_{i_2i_3}|n\rangle-4\langle
n|\hat{H}_{i}\frac{\hat{Q}_n}{\hat{H}_0-\mathcal{E}_n}\hat{H}_{i_1i_2i_3}
|n\rangle+\text{c.c.}\Big]\notag\\
&\quad+12\mathcal{S}_{ii_1i_2i_3}\sum_n\rho_n\Big[ \langle
n|\hat{H}_{i_3}\frac{\hat{Q}_n}{\hat{H}_0-\mathcal{E}_{n}}\delta_n\hat{H}_{i_1i_2}\frac{\hat{Q}_n}{\hat{H}_0-\mathcal{E}_{n}}\hat{H}_i|n\rangle
+\langle
n|\hat{H}_{i_1i_2}\frac{\hat{Q}_n}{\hat{H}_0-\mathcal{E}_{n}}\delta_n\hat{H}_{i}\frac{\hat{Q}_n}{\hat{H}_0-\mathcal{E}_{n}}\hat{H}_{i_3}|n\rangle+\text{c.c.}\Big]\notag\\
&\quad-24\mathcal{S}_{ii_1i_2i_3}\sum_n\rho_n\langle
n|\hat{H}_i\frac{\hat{Q}_n}{\hat{H}_0-\mathcal{E}_{n}}\delta_n\hat{H}_{i_1}\frac{\hat{Q}_n}{\hat{H}_0-\mathcal{E}_{n}}\delta_n\hat{H}_{i_2}\frac{\hat{Q}_n}{\hat{H}_0-\mathcal{E}_{n}}\hat{H}_{i_3}|n\rangle\notag\\
&\quad+24\mathcal{S}_{ii_1i_2i_3}\sum_n\rho_n\langle
n|\hat{H}_i\frac{\hat{Q}_n}{(\hat{H}_0-\mathcal{E}_{n})^2}\hat{H}_{i_1}|n\rangle\langle
n|\hat{H}_{i_2}\frac{\hat{Q}_n}{\hat{H}_0-\mathcal{E}_{n}}\hat{H}_{i_3}|n\rangle\notag\\
&=V\phi_i^{i_1i_2i_3}(\omega_1=0,\omega_2=0,\omega_3=0)=V\mathcal{D}_{i}^{i_1i_2i_3}.\label{D3}
\end{align}
In passing to the last line we set $\omega_1=\omega_2=\omega_3=0$ in Eq.~\eqref{gomega3}. In the derivation, we used
\begin{align}
\mathcal{S}_{i_1i_2i_3}\left(\langle n|n_{i_1i_2i_3}\rangle+3\langle
n_{i_1}|n_{i_2i_3}\rangle\right)+\text{c.c.}=0
\end{align}
and
\begin{align}
\hat{Q}_n|n_{i_1i_2i_3}\rangle&=\mathcal{S}_{i_1i_2i_3}\Big[-\frac{\hat{Q}_n}{\hat{H}_0-\mathcal{E}_n}\hat{H}_{i_1i_2i_3}|n\rangle-3\frac{\hat{Q}_n}{\hat{H}_0-\mathcal{E}_n}\delta_n\hat{H}_{i_1i_2}|n_{i_3}\rangle\notag\\
&\quad\quad\quad\quad\quad\quad\quad-6\frac{\hat{Q}_n}{\hat{H}_0-\mathcal{E}_n}|n_{i_3}\rangle\langle
n|\hat{H}_{i_1}\frac{\hat{Q}_n}{\hat{H}_0-\mathcal{E}_n}\hat{H}_{i_2}|n\rangle-3\frac{\hat{Q}_n}{\hat{H}_0-\mathcal{E}_n}\delta_n\hat{H}_{i_1}|n_{i_2i_3}\rangle\Big],
\end{align}
in addition to the above first- and second-order relations.

\section{tight-binding model}
Let us demonstrate our results with a simple example of a tight-binding
model.  We diagonalize the unperturbed Hamiltonian as
\begin{align}
\hat{H}_0&=\sum_{\bm{k},n}\varepsilon_{\bm{k}n}\hat{\gamma}_{\bm{k}n}^\dagger\hat{\gamma}_{\bm{k}n}.
\end{align}
In this basis, $\hat{H}_{i_1\dots i_N}$ in Eq.~\eqref{defHN2} can be
written as
\begin{align}
\hat{H}_{i_1\dots i_N}&=\sum_{\bm{k},m,n}\hat{\gamma}_{\bm{k}m}^\dagger
h_{i_1\dots i_N}^{\bm{k}mn}\hat{\gamma}_{\bm{k}n}.
\end{align}
The response function $\phi_i^{i_1}(\omega_1)$ and
$\phi_{i}^{i_1i_2}(\omega_1,\omega_2)$ are then given by
\begin{align}
V\phi_i^{i_1}(\omega_1)&=\sum_{\bm{k},n<0}h_{ii_1}^{\bm{k}nn}+\sum_{\bm{k},n<0,m>0}\left(\frac{h_i^{\bm{k}nm}h_{i_1}^{\bm{k}mn}}{\omega_1-(\varepsilon_{\bm{k}m}-\varepsilon_{\bm{k}n})+i\eta}
-\frac{h_{i_1}^{\bm{k}nm}h_i^{\bm{k}mn}}{\omega_1+(\varepsilon_{\bm{k}m}-\varepsilon_{\bm{k}n})+i\eta}\right)\label{tb1}
\end{align}
and
\begin{align}
V\phi_{i}^{i_1i_2}(\omega_1,\omega_2)
&=\sum_{\bm{k},n<0}h_{ii_1i_2}^{\bm{k}nn}+2\mathcal{S}_{i_1i_2}\sum_{\bm{k},n<0,m>0}\left(\frac{h_{ii_1}^{\bm{k}nm}h_{i_2}^{\bm{k}mn}}{\omega_2-(\varepsilon_{\bm{k}m}-\varepsilon_{\bm{k}n})+i\eta}
-\frac{h_{i_2}^{\bm{k}nm}h_{ii_1}^{\bm{k}mn}}{\omega_2+(\varepsilon_{\bm{k}m}-\varepsilon_{\bm{k}n})+i\eta}\right)\notag\\
&\quad+\sum_{\bm{k},n<0,m>0}\left(\frac{h_i^{\bm{k}nm}h_{i_1i_2}^{\bm{k}mn}}{\omega_1+\omega_2-(\varepsilon_{\bm{k}m}-\varepsilon_{\bm{k}n})+2i\eta}
-\frac{h_{i_1i_2}^{\bm{k}nm}h_{i}^{\bm{k}mn}}{\omega_1+\omega_2+(\varepsilon_{\bm{k}m}-\varepsilon_{\bm{k}n})+2i\eta}\right)\notag\\
&\quad+2\mathcal{S}_{i_1i_2}\sum_{\bm{k},n,n'<0,m,m'>0}\frac{h_i^{\bm{k}nm}(h_{i_1}^{\bm{k}mm'}\delta_{n,n'}-h_{i_1}^{\bm{k}n'n}\delta_{m,m'})h_{i_2}^{\bm{k}m'n'}}{\big[\omega_1+\omega_2-(\varepsilon_{\bm{k}m}-\varepsilon_{\bm{k}n})+2i\eta\big]\big[\omega_2-(\varepsilon_{\bm{k}m'}-\varepsilon_{\bm{k}n'})+i\eta\big]}\notag\\
&\quad-2\mathcal{S}_{i_1i_2}\sum_{\bm{k},n,n'<0,m,m'>0}\frac{h_{i_1}^{\bm{k}nm}(h_i^{\bm{k}mm'}\delta_{n,n'}-h_i^{\bm{k}n'n}\delta_{m,m'})h_{i_2}^{\bm{k}m'n'}}{\big[\omega_1+(\varepsilon_{\bm{k}m}-\varepsilon_{\bm{k}n})+i\eta\big]\big[\omega_2-(\varepsilon_{\bm{k}m'}-\varepsilon_{\bm{k}n'})+i\eta\big]}\notag\\
&\quad+2\mathcal{S}_{i_1i_2}\sum_{\bm{k},n,n'<0,m,m'>0}\frac{h_{i_2}^{\bm{k}nm}(h_{i_1}^{\bm{k}mm'}\delta_{n,n'}-h_{i_1}^{\bm{k}n'n}\delta_{m,m'})h_{i}^{\bm{k}m'n'}}{\big[\omega_2+(\varepsilon_{\bm{k}m}-\varepsilon_{\bm{k}n})+i\eta\big]\big[\omega_1+\omega_2+(\varepsilon_{\bm{k}m'}-\varepsilon_{\bm{k}n'})+2i\eta\big]}.\label{tb2}
\end{align}

\begin{figure}[t]
\begin{center}
\includegraphics[width=0.99\textwidth]{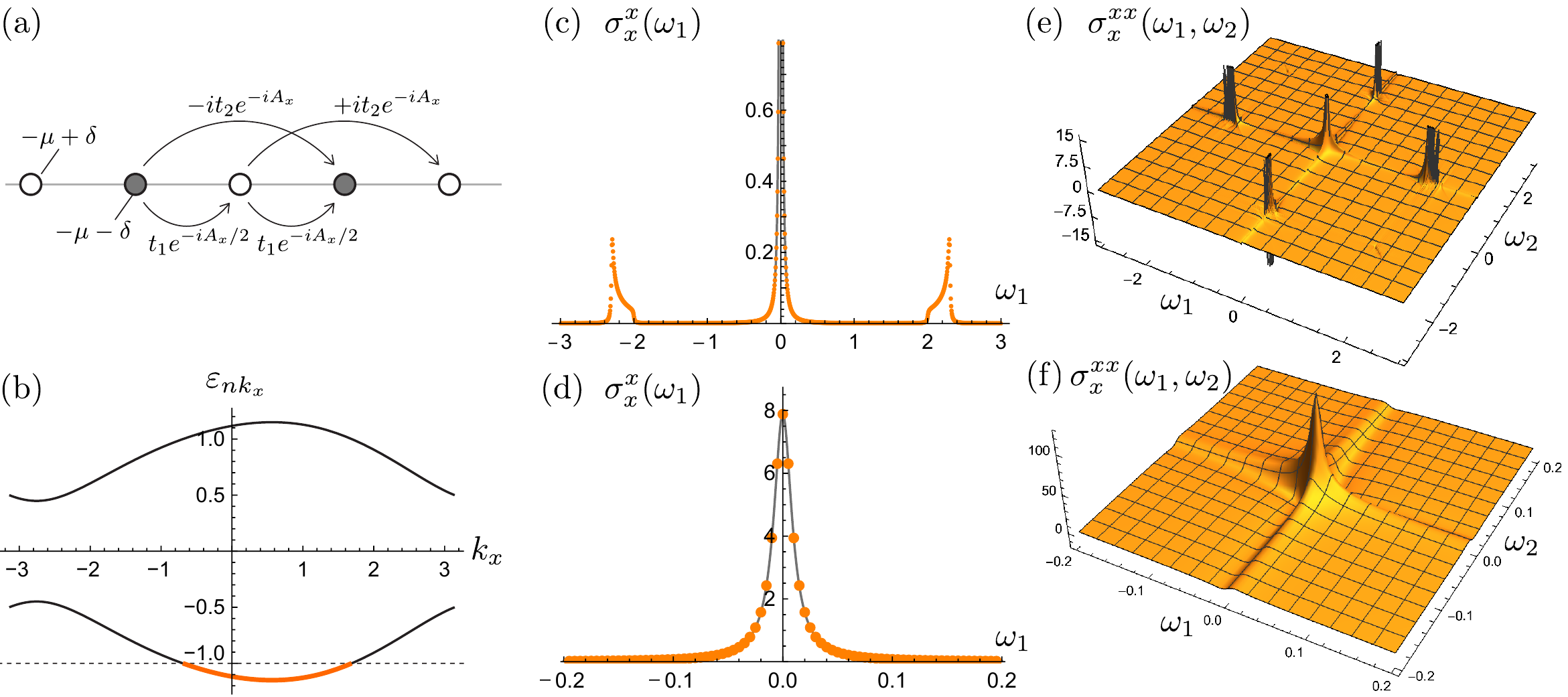} \caption{\label{fig} The
linear and the second-order optical conductivities in the tight-binding
model in Eq.~\eqref{TB}. (a) The real-space illustration of the
model. (b) The band structure $\varepsilon_{nk_x}$ as a function of
$k_x$. The orange part is occupied in the ground state. (c)
$\sigma_x^{x}(\omega_1)$ as a function of $\omega_1\in(-3,3)$. The gray
curve is the fit by Eq.~\eqref{fit}.  (d) The zoom up of (c) for
$\omega_1\in(-0.2,0.2)$. (e) $\sigma_x^{xx}(\omega_1,\omega_2)$ as a
function of $\omega_1, \omega_2\in(-3,3)$. (f) The zoom up of (e). }
\end{center}
\end{figure}

We use the following two-band model~\cite{MorimotoNagaosa2016}, illustrated in Fig.~\ref{fig} (a),
in $d=1$ at zero temperature for the demonstration.
\begin{equation}
H_{k_x}=\begin{pmatrix} -\delta-2t_2\sin(k_x+A_x) &
t_1(e^{iA_x/2}+e^{-i(k_x+A_x/2)})\\ t_1(e^{-iA_x/2}+e^{i(k_x+A_x/2)}) &
\delta+2t_2\sin(k_x+A_x)
\end{pmatrix}.\label{TB}
\end{equation}
This model breaks both the inversion symmetry and the time-reversal
symmetry, resulting in a nonzero $f$-sum for the second-order
conductivity.  We set $t_1=0.5$, $\delta=0.5$, and $t_2=0.125$.  The
crystal momentum $k_x$ takes values $2\pi i_x/L_x$ with $i_x=1,2,\cdots
L_x$ and $L_x=501$.  When the chemical potential is set to be $-1$, the
system is metallic and partially fills the lower band as shown in
Fig.~\ref{fig} (b). The optical conductivities $\sigma_x^{x}(\omega_1)$
and $\sigma_x^{xx}(\omega_1,\omega_2)$ are computed based on
Eqs.~\eqref{gtosigma3}, \eqref{tb1}, and \eqref{tb2} with $\eta=0.01$
for the range of $\omega_1$ and $\omega_2$ in $|\omega_\ell|<10$. The
frequencies are discretized with $\Delta\omega=1/200$.  The obtained
optical conductivities as a function of $\omega$ are shown in
Fig.~\ref{fig} (c) and (e).

\begin{table}[t]
\caption{Numerical results for the tight-binding model in
Eq.~\eqref{TB}. See the main text for the definitions of these
quantities in the actual calculation.}  \label{table}
\begin{center}
\begin{tabular}{ c  c | c  c | c  c | c  c }\hline\hline
\multicolumn{4}{c|}{Linear response $\sigma_x^{x}(\omega_1)$} &
\multicolumn{4}{|c}{Second-order response
$\sigma_x^{xx}(\omega_1,\omega_2)$} \\ \hline \multicolumn{2}{c|}{Drude
weight} & \multicolumn{2}{|c|}{$f$-sum} & \multicolumn{2}{|c|}{Drude
weight} & \multicolumn{2}{|c}{$f$-sum} \\ $\mathcal{D}_x^x$ &
$\frac{1}{L_x}\frac{\partial^2\mathcal{E}_0(A_x)}{\partial A_x^2}$ & $\int
\frac{d\omega_1}{2\pi}\sigma_x^x(\omega_1)$ &
$\frac{1}{2L_x}\langle\frac{\partial^2\hat{H}(A_x)}{\partial
A_x^2}\rangle_0$ & $\mathcal{D}_x^{xx}$ &
$\frac{1}{L_x}\frac{\partial^3\mathcal{E}_0(A_x)}{\partial A_x^3}$ & $\int\int
\frac{d\omega_1d\omega_2}{(2\pi)^2}\sigma_x^{xx}(\omega_1,\omega_2)$ &
$\frac{1}{4L_x}\langle\frac{\partial^3\hat{H}(A_x)}{\partial
A_x^3}\rangle_0$ \\ \hline
\quad$0.0788238$\quad & \quad$0.0788231$\quad & \quad$0.0487034$\quad &
\quad$0.0487345$\quad & \quad$0.0122513$\quad & \quad$0.0122554$\quad &
\quad$0.00594065$\quad & \quad$0.00596566$\quad \\ \hline\hline
\end{tabular}
\end{center}
\end{table}

The Drude weights $\mathcal{D}_x^x$ and $\mathcal{D}_x^{xx}$ are then
determined by fitting $\text{Re}[\sigma_x^{x}(\omega_1)]$ and
$\text{Re}[\sigma_x^{xx}(\omega_1,\omega_2)]$ for small $\omega_\ell$'s
[Fig.~\ref{fig} (d) and (f)] by
\begin{align}
\text{Re}[\sigma_x^{x}(\omega_1)]=\frac{\eta}{\omega^2+\eta^2}\mathcal{D}_x^x,\quad
\text{Re}[\sigma_x^{xx}(\omega_1,\omega_2)]=\frac{\eta^2-\omega_1\omega_2}{(\omega_1^2+\eta^2)(\omega_2^2+\eta^2)}\mathcal{D}_x^{xx}\quad
(|\omega_\ell|\ll1).\label{fit}
\end{align}
For this fit, we use frequencies in the range $|\omega_\ell|<0.2$.  The
Drude weights $\mathcal{D}_x^x$ and $\mathcal{D}_x^{xx}$ obtained this
way are compared with $\partial^2\mathcal{E}_0(A_x)/\partial A_x^2$ and
$\partial^3\mathcal{E}_0(A_x)/\partial A_x^3$ computed separately. As
summarized in Table~\ref{table}, we observe good agreement, confirming
the Kohn formula.

The frequency integration of $\sigma_x^{x}(\omega_1)$ and
$\sigma_x^{xx}(\omega_1,\omega_2)$ in Eq.~\eqref{fsumrule} are
approximated by the Riemannian summation, i.e.,
\begin{align}
\sum_{-10<\omega_1<10}\frac{\Delta\omega}{2\pi}\,\sigma_x^{x}(\omega_1),\quad
\sum_{-10<\omega_1,\omega_2<10}\frac{(\Delta\omega)^2}{(2\pi)^2}\,\sigma_x^{xx}(\omega_1,\omega_2).
\end{align}
The results are compared with
$\langle\partial^2\hat{H}(A_x)/\partial A_x^2\rangle_0$ and
$\langle\partial^3\hat{H}(A_x)/\partial A_x^3\rangle_0$ computed
separately. Again, we find that they agree well, verifying the $f$-sum
rule.

\section{Conclusion}
In this work, we studied the non-linear conductivity $\sigma_i^{i_1 \dots i_N}(t_1,\dots,t_N)$
with respect to the spatially uniform electric field. We provided the detailed discussion on the relation between the instantaneous response and the $f$-sum rule, hinted in Ref.~\cite{2003.10390}.
We also derived explicit expressions of the
nonlinear optical conductivity $\sigma_i^{i_1 \dots i_N}(\omega_1,\dots,\omega_N)$ of the orders $N=2$ and $3$ for general
quantum many-body systems, in terms of correlation functions.  Based on
the explicit formulas obtained, we proved the nonlinear generalizations
of the Kohn formula to the second- and third-order
optical conductivity, confirming the results in Ref.~\cite{2003.10390} derived
in a general framework but without explicit expressions of the non-linear conductivities.
The obtained non-linear $f$-sum rules and the Kohn formulas are valid for any finite size,
and thus also could be used in the thermodynamic limit with an appropriate procedure.
The exact $f$-sum rules for the finite size systems would be useful for the benchmarking of numerical calculations.

As a demonstration, we applied our formulation to a simple tight-binding model in
one dimension, and numerically verified the non-linear $f$-sum rule and Drude weight of
the second order. While this demonstration was done for the non-interacting
electrons, we emphasize that our formulation and results of the present paper is
applicable to very general quantum many-body systems with interactions.

Since our discussion did not assume the translation invariance, both the generalized $f$-sum rules and the Kohn formulas should be valid even in the presence of disorder.
For example, in the localized phases, all energy levels are insensitive to the the applied vector potential because the effect of vector potential can be converted to the twist of the boundary condition by a local gauge transformation. (In other words, only extended states can be affected by the vector potential.)  This is consistent with our expectation that both the linear and nonlinear Drude weight should vanish in the localized phases. This discussion suggests the presence of residual Drude peak in weakly disordered phases. We leave the detailed investigation of such possibility as a future work.

\begin{acknowledgments}
H.W. thanks Takahiro Morimoto for useful discussions.
The work of M.O. was supported in part by MEXT/JSPS KAKENHI Grant Nos. JP19H01808 and JP17H06462, and JST CREST Grant Number JPMJCR19T2, Japan.
The work of H.W. is supported by JST PRESTO Grant No. JPMJPR18LA.
\end{acknowledgments}

\bibliography{bibs}

\appendix

\section{Expression of $\phi_{i}^{i_1i_2i_3}$}
\label{phithird}

Here we present the expression of $\phi_{i}^{i_1i_2i_3}(t_1,t_2,t_3)$ and $\phi_{i}^{i_1i_2i_3}(\omega_1,\omega_2,\omega_3)$.  In the presence of the time-translation symmetry, Eq.~\eqref{gtt3} becomes
\begin{align}
&V\phi_{i}^{i_1i_2i_3}(t_1,t_2,t_3)\notag\\
&=\delta(t_1)\delta(t_2)\delta(t_3)\langle\hat{H}_{i i_1 i_2
i_3}\rangle_0\notag\\
&\quad+3\mathcal{S}_{i_1i_2i_3}\delta(t_1)\delta(t_2)\theta(t_3)(-i)\langle[\hat{H}_{ii_1i_2},e^{-i\hat{H}_0t_3}\hat{H}_{i_3}e^{i\hat{H}_0t_3}]\rangle_0\notag\\
&\quad+3\mathcal{S}_{i_1i_2i_3}\delta(t_1)\delta(t_3-t_2)\theta(t_2)(-i)\langle[\hat{H}_{i
i_1},e^{-i\hat{H}_0t_2}\hat{H}_{i_2i_3}e^{i\hat{H}_0t_2}]\rangle_0\notag\\
&\quad+6\mathcal{S}_{i_1i_2i_3}\delta(t_1)\theta(t_2)\theta(t_3-t_2)(-i)^2\langle[[\hat{H}_{i
i_1},e^{-i\hat{H}_0t_2}\hat{H}_{i_2}e^{i\hat{H}_0t_2}],e^{-i\hat{H}_0t_3}\hat{H}_{i_3}e^{i\hat{H}_0t_3}]\rangle_0\notag\\
&\quad+\delta(t_2-t_1)\delta(t_3-t_2)\theta(t_1)(-i)\langle[\hat{H}_{i},e^{-i\hat{H}_0t_1}\hat{H}_{i_1i_2i_3}e^{i\hat{H}_0t_1}]\rangle_0\notag\\
&\quad+3\mathcal{S}_{i_1i_2i_3}\delta(t_2-t_1)\theta(t_1)\theta(t_3-t_1)(-i)^2\langle[[\hat{H}_{i},e^{-i\hat{H}_0t_1}\hat{H}_{i_1i_2}e^{i\hat{H}_0t_1}],e^{-i\hat{H}_0t_3}\hat{H}_{i_3}e^{i\hat{H}_0t_3}]\rangle_0\notag\\
&\quad+3\mathcal{S}_{i_1i_2i_3}\delta(t_3-t_2)\theta(t_1)\theta(t_2-t_1)(-i)^2\langle[[\hat{H}_{i},e^{-i\hat{H}_0t_1}\hat{H}_{i_1}e^{i\hat{H}_0t_1}],e^{-i\hat{H}_0t_2}\hat{H}_{i_2i_3}e^{i\hat{H}_0t_2}]\rangle_0\notag\\
&\quad+6\mathcal{S}_{i_1i_2i_3}\theta(t_1)\theta(t_2-t_1)\theta(t_3-t_2)(-i)^3\langle[[[\hat{H}_{i},e^{-i\hat{H}_0t_1}\hat{H}_{i_1}e^{i\hat{H}_0t_1}],e^{-i\hat{H}_0t_2}\hat{H}_{i_2}e^{i\hat{H}_0t_2}],e^{-i\hat{H}_0t_3}\hat{H}_{i_3}e^{i\hat{H}_0t_3}]\rangle_0.\label{gt3}
\end{align}

Fourier transformation of Eq.~\eqref{gt3} gives
\begin{align}
&V\phi_{i}^{i_1i_2i_3}(\omega_1,\omega_2,\omega_3)=\langle\hat{H}_{i i_1
i_2 i_3}\rangle_0+I_1+I_2+I_3,\label{gomega3}
\end{align}
where
\begin{align}
&I_1=3\mathcal{S}_{i_1i_2i_3}\sum_n\rho_n\langle
n|\Big(\delta_n\hat{H}_{ii_1i_2}\frac{\hat{Q}_n}{\omega_3-\hat{H}_0+\mathcal{E}_n+i\eta}\delta_n\hat{H}_{i_3}-\delta_n\hat{H}_{i_3}\frac{\hat{Q}_n}{\omega_3+\hat{H}_0-\mathcal{E}_n+i\eta}\delta_n\hat{H}_{ii_1i_2}\Big)|n\rangle\notag\\
&+3\mathcal{S}_{i_1i_2i_3}\sum_n\rho_n\langle
n|\Big(\delta_n\hat{H}_{ii_1}\frac{\hat{Q}_n}{\omega_2+\omega_3-\hat{H}_0+\mathcal{E}_n+2i\eta}\delta_n\hat{H}_{i_2i_3}-\delta_n\hat{H}_{i_2i_3}\frac{\hat{Q}_n}{\omega_2+\omega_3+\hat{H}_0-\mathcal{E}_n+2i\eta}\delta_n\hat{H}_{ii_1}\Big)|n\rangle\notag\\
&+6\mathcal{S}_{i_1i_2i_3}\sum_n\rho_n \Big[\langle
n|\delta_n\hat{H}_{ii_1}\frac{\hat{Q}_n}{\omega_2+\omega_3-\hat{H}_0+\mathcal{E}_{n}+2i\eta}\delta_n\hat{H}_{i_2}\frac{\hat{Q}_n}{\omega_3-\hat{H}_0+\mathcal{E}_{n}+i\eta}\delta_n\hat{H}_{i_3}|n\rangle\notag\\
&\quad\quad\quad\quad\quad\quad\quad\quad-\langle
n|\delta_n\hat{H}_{i_2}\frac{\hat{Q}_n}{\omega_2+\hat{H}_0-\mathcal{E}_{n}+i\eta}\delta_n\hat{H}_{ii_1}\frac{\hat{Q}_n}{\omega_3-\hat{H}_0+\mathcal{E}_{n}+i\eta}\delta_n\hat{H}_{i_3}|n\rangle\notag\\
&\quad\quad\quad\quad\quad\quad\quad\quad+\langle
n|\delta_n\hat{H}_{i_3}\frac{\hat{Q}_n}{\omega_3+\hat{H}_0-\mathcal{E}_{n}+i\eta}\delta_n\hat{H}_{i_2}\frac{\hat{Q}_n}{\omega_2+\omega_3+\hat{H}_0-\mathcal{E}_{n}+2i\eta}\delta_n\hat{H}_{ii_1}|n\rangle\Big]\notag\\
&+\sum_n\rho_n\langle
n|\Big(\delta_n\hat{H}_{i}\frac{\hat{Q}_n}{\omega_1+\omega_2+\omega_3-\hat{H}_0+\mathcal{E}_n+3i\eta}\delta_n\hat{H}_{i_1i_2i_3}-\delta_n\hat{H}_{i_1i_2i_3}\frac{\hat{Q}_n}{\omega_1+\omega_2+\omega_3+\hat{H}_0-\mathcal{E}_n+3i\eta}\delta_n\hat{H}_{i}\Big)
|n\rangle,
\end{align}
\begin{align}
&I_2=3\mathcal{S}_{i_1i_2i_3}\sum_n\rho_n \Big[\langle
n|\delta_n\hat{H}_{i}\frac{\hat{Q}_n}{\omega_1+\omega_2+\omega_3-\hat{H}_0+\mathcal{E}_{n}+3i\eta}\delta_n\hat{H}_{i_1i_2}\frac{\hat{Q}_n}{\omega_3-\hat{H}_0+\mathcal{E}_{n}+i\eta}\delta_n\hat{H}_{i_3}|n\rangle\notag\\
&\quad\quad\quad\quad\quad\quad\quad\quad-\langle
n|\delta_n\hat{H}_{i_1i_2}\frac{\hat{Q}_n}{\omega_1+\omega_2+\hat{H}_0-\mathcal{E}_{n}+2i\eta}\delta_n\hat{H}_{i}\frac{\hat{Q}_n}{\omega_3-\hat{H}_0+\mathcal{E}_{n}+i\eta}\delta_n\hat{H}_{i_3}|n\rangle\notag\\
&\quad\quad\quad\quad\quad\quad\quad\quad+\langle
n|\delta_n\hat{H}_{i_3}\frac{\hat{Q}_n}{\omega_3+\hat{H}_0-\mathcal{E}_{n}+i\eta}\delta_n\hat{H}_{i_1i_2}\frac{\hat{Q}_n}{\omega_1+\omega_2+\omega_3+\hat{H}_0-\mathcal{E}_{n}+3i\eta}\delta_n\hat{H}_{i}|n\rangle\Big]\notag\\
&+3\mathcal{S}_{i_1i_2i_3}\sum_n\rho_n \Big[\langle
n|\delta_n\hat{H}_{i}\frac{\hat{Q}_n}{\omega_1+\omega_2+\omega_3-\hat{H}_0+\mathcal{E}_{n}+2i\eta}\delta_n\hat{H}_{i_1}\frac{\hat{Q}_n}{\omega_2+\omega_3-\hat{H}_0+\mathcal{E}_{n}+i\eta}\delta_n\hat{H}_{i_2i_3}|n\rangle\notag\\
&\quad\quad\quad\quad\quad\quad\quad\quad-\langle
n|\delta_n\hat{H}_{i_1}\frac{\hat{Q}_n}{\omega_1+\hat{H}_0-\mathcal{E}_{n}+i\eta}\delta_n\hat{H}_{i}\frac{\hat{Q}_n}{\omega_2+\omega_3-\hat{H}_0+\mathcal{E}_{n}+i\eta}\delta_n\hat{H}_{i_2i_3}|n\rangle\notag\\
&\quad\quad\quad\quad\quad\quad\quad\quad+\langle
n|\delta_n\hat{H}_{i_2i_3}\frac{\hat{Q}_n}{\omega_2+\omega_3+\hat{H}_0-\mathcal{E}_{n}+i\eta}\delta_n\hat{H}_{i_1}\frac{\hat{Q}_n}{\omega_1+\omega_2+\omega_3+\hat{H}_0-\mathcal{E}_{n}+2i\eta}\delta_n\hat{H}_{i}|n\rangle\Big],
\end{align}
and
\begin{align}
&I_3=6\mathcal{S}_{i_1i_2i_3}\sum_n\rho_n\langle
n|\delta_n\hat{H}_i\frac{\hat{Q}_n}{\omega_1+\omega_2+\omega_3-\hat{H}_0+\mathcal{E}_{n}+3i\eta}\delta_n\hat{H}_{i_1}\frac{\hat{Q}_n}{\omega_2+\omega_3-\hat{H}_0+\mathcal{E}_{n}+2i\eta}\delta_n\hat{H}_{i_2}\frac{\hat{Q}_n}{\omega_3-\hat{H}_0+\mathcal{E}_{n}+i\eta}\delta_n\hat{H}_{i_3}|n\rangle\notag\\
&-6\mathcal{S}_{i_1i_2i_3}\sum_n\rho_n\langle
n|\delta_n\hat{H}_{i_1}\frac{\hat{Q}_n}{\omega_1+\hat{H}_0-\mathcal{E}_{n}+i\eta}\delta_n\hat{H}_i\frac{\hat{Q}_n}{\omega_2+\omega_3-\hat{H}_0+\mathcal{E}_{n}+2i\eta}\delta_n\hat{H}_{i_2}\frac{\hat{Q}_n}{\omega_3-\hat{H}_0+\mathcal{E}_{n}+i\eta}\delta_n\hat{H}_{i_3}|n\rangle\notag\\
&+6\mathcal{S}_{i_1i_2i_3}\sum_n\rho_n\langle
n|\delta_n\hat{H}_{i_3}\frac{\hat{Q}_n}{\omega_3+\hat{H}_0-\mathcal{E}_{n}+i\eta}\delta_n\hat{H}_{i_2}\frac{\hat{Q}_n}{\omega_2+\omega_3+\hat{H}_0-\mathcal{E}_{n}+2i\eta}\delta_n\hat{H}_i\frac{\hat{Q}_n}{\omega_1-\hat{H}_0+\mathcal{E}_{n}+i\eta}\delta_n\hat{H}_{i_1}|n\rangle\notag\\
&-6\mathcal{S}_{i_1i_2i_3}\sum_n\rho_n\langle
n|\delta_n\hat{H}_{i_3}\frac{\hat{Q}_n}{\omega_3+\hat{H}_0-\mathcal{E}_{n}+i\eta}\delta_n\hat{H}_{i_2}\frac{\hat{Q}_n}{\omega_2+\omega_3+\hat{H}_0-\mathcal{E}_{n}+2i\eta}\delta_n\hat{H}_{i_1}\frac{\hat{Q}_n}{\omega_1+\omega_2+\omega_3+\hat{H}_0-\mathcal{E}_{n}+3i\eta}\delta_n\hat{H}_i|n\rangle\notag\\
&+3\mathcal{S}_{i_1i_2i_3}\sum_n\rho_n\langle
n|\delta_n\hat{H}_i\frac{\hat{Q}_n}{\omega_1+\omega_2+\omega_3-\hat{H}_0+\mathcal{E}_{n}+3i\eta}\delta_n\hat{H}_{i_1}|n\rangle\langle
n|\delta_n\hat{H}_{i_2}\frac{\hat{Q}_n}{(\omega_3-\hat{H}_0+\mathcal{E}_{n}+i\eta)(\omega_2+\hat{H}_0-\mathcal{E}_{n}+i\eta)}\delta_n\hat{H}_{i_3}|n\rangle\notag\\
&+3\mathcal{S}_{i_1i_2i_3}\sum_n\rho_n\langle
n|\delta_n\hat{H}_i\frac{\hat{Q}_n}{\omega_1-\hat{H}_0+\mathcal{E}_{n}+i\eta}\delta_n\hat{H}_{i_1}|n\rangle\langle
n|\delta_n\hat{H}_{i_2}\frac{\hat{Q}_n}{(\omega_3-\hat{H}_0+\mathcal{E}_{n}+i\eta)(\omega_2+\hat{H}_0-\mathcal{E}_{n}+i\eta)}\delta_n\hat{H}_{i_3}|n\rangle\notag\\
&-3\mathcal{S}_{i_1i_2i_3}\sum_n\rho_n\langle
n|\delta_n\hat{H}_{i_3}\frac{\hat{Q}_n}{(\omega_2-\hat{H}_0+\mathcal{E}_{n}+i\eta)(\omega_3+\hat{H}_0-\mathcal{E}_{n}+i\eta)}\delta_n\hat{H}_{i_2}|n\rangle\langle
n|\delta_n\hat{H}_{i_1}\frac{\hat{Q}_n}{\omega_1+\hat{H}_0-\mathcal{E}_{n}+i\eta}\delta_n\hat{H}_i|n\rangle\notag\\
&-3\mathcal{S}_{i_1i_2i_3}\sum_n\rho_n\langle
n|\delta_n\hat{H}_{i_3}\frac{\hat{Q}_n}{(\omega_2-\hat{H}_0+\mathcal{E}_{n}+i\eta)(\omega_3+\hat{H}_0-\mathcal{E}_{n}+i\eta)}\delta_n\hat{H}_{i_2}|n\rangle\langle
n|\delta_n\hat{H}_{i_1}\frac{\hat{Q}_n}{\omega_1+\omega_2+\omega_3+\hat{H}_0-\mathcal{E}_{n}+3i\eta}\delta_n\hat{H}_i|n\rangle\notag\\
&-6\mathcal{S}_{i_1i_2i_3}\sum_n\rho_n\langle
n|\delta_n\hat{H}_i\frac{\hat{Q}_n}{(\omega_1-\hat{H}_0+\mathcal{E}_{n}+i\eta)(\omega_1+\omega_2+\omega_3-\hat{H}_0+\mathcal{E}_{n}+3i\eta)}\delta_n\hat{H}_{i_1}|n\rangle\notag\\
&\quad\quad\quad\quad\quad\quad\quad\times\langle
n|\delta_n\hat{H}_{i_2}\frac{\hat{Q}_n(\hat{H}_0-\mathcal{E}_{n})}{(\omega_3-\hat{H}_0+\mathcal{E}_{n}+i\eta)(\omega_2+\hat{H}_0-\mathcal{E}_{n}+i\eta)}\delta_n\hat{H}_{i_3}|n\rangle\notag\\
&-6\mathcal{S}_{i_1i_2i_3}\sum_n\rho_n\langle
n|\delta_n\hat{H}_{i_3}\frac{\hat{Q}_n(\hat{H}_0-\mathcal{E}_{n})}{(\omega_2-\hat{H}_0+\mathcal{E}_{n}+i\eta)(\omega_3+\hat{H}_0-\mathcal{E}_{n}+i\eta)}\delta_n\hat{H}_{i_2}|n\rangle\notag\\
&\quad\quad\quad\quad\quad\quad\quad\times\langle
n|\delta_n\hat{H}_{i_1}\frac{\hat{Q}_n}{(\omega_1+\omega_2+\omega_3+\hat{H}_0-\mathcal{E}_{n}+3i\eta)(\omega_1+\hat{H}_0-\mathcal{E}_{n}+i\eta)}\delta_n\hat{H}_i|n\rangle.
\end{align}

\end{document}